\documentclass[10pt]{iopart}
\usepackage{graphicx}
\usepackage{iopams}

\newcommand{\half}{{\textstyle \frac{1}{2}}}

\def\hbarit {{\mathchar'26\mkern-11muh}} 
\newcommand{\bfsfI}{\mbox{\sffamily\bfseries{I}}}
\newcommand{\bfsfT}{\mbox{\sffamily\bfseries{T}}}
\newcommand{\bfsfS}{\mbox{\sffamily\bfseries{S}}}
\newcommand{\bfsfF}{\mbox{\sffamily\bfseries{F}}}
\newcommand{\bfsfG}{\mbox{\sffamily\bfseries{G}}}
\newcommand{\bfsfH}{\mbox{\sffamily\bfseries{H}}}
\newcommand{\bfsfM}{\mbox{\sffamily\bfseries{M}}}
\newcommand{\bfsfU}{\mbox{\sffamily\bfseries{U}}}
\newcommand{\bfsff}{\mbox{\sffamily\bfseries{f}}}
\newcommand{\bfsfg}{\mbox{\sffamily\bfseries{g}}}
\newcommand{\bfsfh}{\mbox{\sffamily\bfseries{h}}}
\newcommand{\bfsfs}{\mbox{\sffamily\bfseries{s}}}
\newcommand{\bfsfu}{\mbox{\sffamily\bfseries{u}}}
\begin{document}
\jl{1} 

\title[Field quantization in inhomogeneous dielectrics with spatio-temporal
  dispersion]{Field quantization in inhomogeneous anisotropic dielectrics
  with spatio-temporal dispersion}

\author{L G Suttorp}

\address{Instituut voor Theoretische Fysica, Universiteit van Amsterdam,
Valckenierstraat 65, 1018 XE Amsterdam, The Netherlands}

\begin{abstract}
A quantum damped-polariton model is constructed for an inhomogeneous
anisotropic linear dielectric with arbitrary dispersion in space and
time. The model Hamiltonian is completely diagonalized by determining the
creation and annihilation operators for the fundamental polariton modes as
specific linear combinations of the basic dynamical variables. Explicit
expressions are derived for the time-dependent operators describing the
electromagnetic field, the dielectric polarization and the noise term in
the latter. It is shown how to identify bath variables that generate the
dissipative dynamics of the medium.

\end{abstract}   


\submitted 

\maketitle

\section{Introduction}\label{sec1}
Quantization of the electromagnetic field in a linear dielectric medium is
a nontrivial task for various reasons. First of all, since the response of
a dielectric to external fields is frequency-dependent in general, temporal
dispersion should be taken into account. The well-known Kramers-Kronig
relation implies that dispersion is necessarily accompanied by dissipation,
so that the quantization procedure has to describe an electromagnetic field
that is subject to damping.  Furthermore, since the transverse and the
longitudinal parts of the electromagnetic field play a different role in
the dynamics, the quantization scheme should treat these parts
separately. For inhomogeneous and spatially dispersive media this leads to
complications in the quantization procedure, which further increase in the
presence of anisotropy.

When the losses in a specific range of frequencies are small, temporal
dispersion can be neglected. Field quantization in an inhomogeneous
isotropic dielectric medium without spatio-temporal dispersion has been
accomplished by employing a generalized transverse gauge, which depends on
the dielectric constant \cite{KVW87}--\cite{WSL03}.

A phenomenological scheme for field quantization in lossy dielectrics has
been formulated on the basis of the fluctuation-dissipation theorem
\cite{GW95}--\cite{KSW01}. By adding a fluctuating noise term to the
Maxwell equations and postulating specific commutation relations for the
operator associated with the noise, one arrives at a quantization procedure that
has been quite successful in describing the electromagnetic field in lossy
dielectrics. An equivalent description in terms of auxiliary fields has
been given as well \cite{T98,TKSW01}, while a related formalism has been
presented recently \cite{FS05}. However, all of these quantization schemes
have the drawback that the precise physical nature of the noise term is not
obvious, since its connection to the basic dynamical variables of the system
is left unspecified. As a consequence, the status of the commutation
relations for the noise operator is that of a postulate.
 
A justification of the above phenomenological quantization scheme has been
sought by adopting a suitable model for lossy dielectrics. To that end use
has been made of an extended version of the Hopfield polariton model
\cite{H58} in which damping effects are accounted for by adding a dynamical
coupling to a bath environment. Huttner and Barnett \cite{HB92a,HB92b} were
the first to employ such a damped-polariton model in order to achieve field
quantization for a lossy dielectric. Their treatment, which is confined to
a spatially homogeneous medium, yields an explicit expression for the noise
term as a linear combination of the canonical variables of the model. In a
later development, an alternative formulation of the quantization procedure
in terms of path integrals has been given \cite{B99}, while Laplace
transformations have been used to simplify the original formalism
\cite{WS01}.  More recently, the effects of spatial inhomogeneities in the
medium have been incorporated by solving an inhomogeneous version of the
damped-polariton model \cite{SWo04}--\cite{WS04}.

In this way a full understanding of the phenomenological quantization
scheme has been reached, at least for those dielectrics that can be
represented by the damped-polariton models mentioned above. The latter
proviso implies a limitation in various ways. First, one would like to
include in a general model not only the effects of spatial inhomogeneity,
but also those of spatial dispersion. Furthermore, it would be desirable to
incorporate the consequences of spatial anisotropy, so that the theory
encompasses crystalline media as well. Finally, while treating temporal
dispersion and the associated damping, we would like to refrain from
introducing a bath environment in the Hamiltonian from the start. Instead,
we wish to formulate the Hamiltonian in terms of a full set of material
variables, from which the dielectric polarization emerges by a suitable
projection.  In this way we will be able to account for any temporal
dispersion that is compatible with a few fundamental principles like
causality and net dielectric loss. For a homogeneous isotropic dielectric
without spatial dispersion such an approach has been suggested before
\cite{HB92b,DF98}.

Recently, several attempts have been made to remove some of the limitations
that are inherent to the earlier treatments. In \cite{B06} the effects of
spatial dispersion are considered in a path-integral formalism for a model
that is a generalization of that of the original Huttner-Barnett approach. The
discussion is confined to homogeneous dielectrics and to leading orders in
the wavenumber, so that an analysis of the effects of arbitrary spatial
dispersion in an inhomogeneous medium is out of reach. In \cite{PL05}
crystalline media have been discussed in the framework of a
damped-polariton model with an anisotropic tensorial bath coupling. A
complete diagonalization of the model along the lines of \cite{HB92a,HB92b}
turned out to face difficulties due to the tensorial complexity, so that
the full dynamics of the model is not presented. Both spatial dispersion
and anisotropy are incorporated in the quantization scheme discussed in
\cite{SSG01}. Use is made of a Langevin approach in which a damping term of
a specific form is introduced. The commutation relations for the noise
operator are postulated, as in the phenomenological quantization
scheme. Finally, several treatments have appeared in which a dielectric
model is formulated while avoiding the explicit introduction of a bath
\cite{KA04,BS06}. However, a complete expression for the noise polarization
operator in terms of the basic dynamical variables of the model is not
presented in these papers. A direct proof of the algebraic properties of
the latter operator is not furnished either.

In the present paper, we shall show how the damped-polariton model can be
generalized in such a way that all of the above restrictions are
removed. As we shall see, our general model describes the quantization and
the time evolution of the electromagnetic field in an inhomogeneous
anisotropic lossy dielectric with arbitrary spatio-temporal dispersion. A
crucial step in arriving at our goals will be the complete diagonalization
of the Hamiltonian. It will lead to explicit expressions for the operators
describing the electromagnetic field and the dielectric polarization, and
for the noise contribution contained in the latter. In this way the
commutation relations for the noise operator will be derived rigorously
from our general model, instead of being postulated along the lines of the
phenomenological scheme. Finally, we shall make contact with previous
treatments by showing how to construct a bath that generates 
damping phenomena in the dynamical evolution of the model.

\section{Model Hamiltonian}\label{sec2}
In this section we shall construct the general form of the Hamiltonian for
a polariton model describing an anisotropic inhomogeneous dispersive
dielectric. The result, which we shall obtain by starting from a few
general principles, will contain several coefficients that can be chosen at
will. As we shall see in a subsequent section, these coefficients can be
adjusted in such a way that the susceptibility gets the appropriate form
for any causal lossy dielectric that we would like to describe.

The Hamiltonian of the electromagnetic field is taken to have the standard form:
\begin{equation}
H_{f}=\int d{\bf r} \left\{\frac{1}{2\varepsilon_0}
[\bPi({\bf r})]^2+\frac{1}{2\mu_0}[\bnabla\wedge{\bf A}({\bf r})]^2\right\}
\label{2.1}
\end{equation}
with the Hermitian vector potential ${\bf A}({\bf r})$ and its associated Hermitian
canonical momentum ${\bPi}({\bf r})$. We use the Coulomb gauge $\bnabla\cdot{\bf
A}=0$. In this gauge both ${\bPi}$ and ${\bf A}$ are transverse. The
canonical commutation relations read
\begin{equation}
\fl [\bPi({\bf r}), {\bf A}({\bf r}')]=-\rmi\, \hbarit\, 
\bdelta_T({\bf r}-{\bf  r}')\quad , \quad
[\bPi({\bf r}), \bPi({\bf r}')]=0 \quad , \quad 
[{\bf A}({\bf r}), {\bf A}({\bf r}')]=0
\label{2.2}
\end{equation}
where the transverse delta function is defined as $\bdelta_T({\bf
r})=\bfsfI\, \delta({\bf r})+\bnabla\bnabla(4\pi r)^{-1}$, with $\bfsfI$
the unit tensor.

The Hamiltonian of the dielectric material medium is supposed to have the
general form
\begin{equation}
H_{m}=\hbarit \int d{\bf r}\int_0^\infty d\omega \; \omega \, \,
{\bf  C}_m^{\dagger}({\bf r},\omega)\cdot {\bf  C}_m({\bf r},\omega)
\label{2.3}
\end{equation}
with the standard commutation relations for the creation and annihilation
operators:
\begin{equation}
\fl [ {\bf  C}_m({\bf r},\omega),{\bf  C}_m^{\dagger}({\bf r}',\omega')]=\bfsfI \,
  \delta({\bf r}-{\bf r}')\, \delta(\omega-\omega')\quad , \quad
[ {\bf  C}_m({\bf r},\omega),{\bf  C}_m({\bf r}',\omega')]=0.
\label{2.4}
\end{equation}
The medium operators commute with the field operators.

The material creation and annihilation operators are assumed to form a
complete set describing all material degrees of freedom. Hence, any
material dynamical variable, for instance the dielectric polarization
density, can be expressed in terms of these operators. For a linear
dielectric medium, the Hermitian polarization density is a linear
combination of the medium operators, which has the general form:
\begin{equation}
{\bf P}({\bf r})=- \rmi \, \hbarit \int d{\bf r}'\int_0^\infty d\omega'\,
{\bf  C}_m({\bf r}',\omega')\cdot \bfsfT({\bf r}',{\bf r},\omega')+{\rm h.c.}
\label{2.5}
\end{equation}
The complex tensorial coefficient $\bfsfT$ appearing in this expression
will be determined later on, when the dielectric susceptibility is properly
identified. On a par with ${\bf P}$ we define its associated canonical
momentum density ${\bf W}$, again as a linear combination of the medium
operators
\begin{equation}
{\bf W}({\bf r})=- \int d{\bf r}'\int_0^\infty d\omega'\,\omega'\, 
{\bf  C}_m({\bf r}',\omega')\cdot \bfsfS({\bf r}',{\bf r},\omega')
+{\rm h.c.}
\label{2.6}
\end{equation}
with a new complex tensorial coefficient $\bfsfS$ that is closely related to
$\bfsfT$, as we shall see below. For future convenience we
inserted a factor $\omega'$ in the integrand and a minus sign in front of
the integral.

As ${\bf W}$ and ${\bf P}$ are a canonical pair, they must satisfy the
standard commutation relations
\begin{equation}
\fl [{\bf W}({\bf r}),{\bf P}({\bf r}')]=-\rmi\, \hbarit\,\bfsfI\, \delta({\bf
  r}-{\bf r}') \quad , \quad [{\bf W}({\bf r}),{\bf W}({\bf r}')]=0 \quad ,
  \quad [{\bf P}({\bf r}),{\bf P}({\bf r}')]=0.
\label{2.7}
\end{equation}
Hence, the coefficients $\bfsfS$ and $\bfsfT$ have to fulfill the
requirements:
\begin{eqnarray}
\int d{\bf r}''\int_0^\infty d\omega'' \, 
\tilde{\bfsfT}({\bf r}'',{\bf r},\omega'')\cdot
{\bfsfT}^\ast({\bf r}'',{\bf r}',\omega'')-{\rm c.c.}=0
\label{2.8}\\
\int d{\bf r}''\int_0^\infty d\omega'' \, \omega''\,
\tilde{\bfsfS}({\bf r}'',{\bf r},\omega'')\cdot
{\bfsfT}^\ast({\bf r}'',{\bf r}',\omega'')+{\rm c.c.}=
\bfsfI \,  \delta({\bf r}-{\bf r}')
\label{2.9}\\
\int d{\bf r}''\int_0^\infty d\omega'' \, \omega''^2\, 
\tilde{\bfsfS}({\bf r}'',{\bf r},\omega'')\cdot
{\bfsfS}^\ast({\bf r}'',{\bf r}',\omega'')-{\rm c.c.}=0
\label{2.10}
\end{eqnarray}
where the tilde denotes the transpose of a tensor and the asterisk the
complex conjugate.

Furthermore, the Hamiltonian should contain terms describing the interaction
between the field and the medium. Two contributions can be distinguished: a
transverse part and a longitudinal part. In a minimal-coupling scheme,
which we shall adopt here, the transverse part is a bilinear expression
involving the transverse vector potential ${\bf A}$ and the canonical
momentum density ${\bf W}$. To ensure compatibility with Maxwell's
equations an expression quadratic in ${\bf A}$ should be present as well,
as we shall see in the following. For dielectrics with spatial dispersion
both expressions are non-local. The general form of the transverse
contribution to the interaction Hamiltonian is
\begin{equation}
\fl H_{i}=-\hbarit \int d{\bf r}\int d{\bf r}'\, {\bf W}({\bf r})\cdot 
\bfsfF_1({\bf r},{\bf r}')\cdot{\bf A}({\bf r}')+
\half\, \hbarit \int d{\bf r}\int d{\bf r}'\, {\bf A}({\bf r})\cdot 
{\bfsfF}_2({\bf r},{\bf r}')\cdot{\bf A}({\bf r}')
\label{2.11}
\end{equation}
with real tensorial coefficients $\bfsfF_1$ and $\bfsfF_2$ that will be fixed
in due time. In view of the form of the second term we may take 
$\bfsfF_2$ to be symmetric upon interchanging both its spatial variables and its
indices. 

The longitudinal contribution to the interaction Hamiltonian is
given by the electrostatic energy involving the polarization density, which
reads
\begin{equation}
H_{\rm es}=\frac{1}{2\varepsilon_0}\int d{\bf r}\, \left\{[{\bf P}({\bf
    r})]_L\right\}^2=
\int d{\bf r}\int d{\bf r}' \, 
\frac{\bnabla\cdot{\bf P}({\bf r})\, \bnabla'\cdot
{\bf P}({\bf r}')}{8\pi\varepsilon_0|{\bf r}-{\bf r}'|}.
\label{2.12}
\end{equation}
Here the subscript $L$ denotes the longitudinal part of the polarization
density, which is defined as $[{\bf P}({\bf r})]_L=-\bnabla\int d{\bf r}'
{\bf P}({\bf r}')\cdot\bnabla(4\,\pi\, |{\bf r}-{\bf r}'|)^{-1}$. Furthermore,
$\bnabla'$ is the spatial derivative acting on a function of ${\bf r}'$.

The total Hamiltonian $H=H_f+H_m+H_i+H_{\rm es}$ is given by the sum of
(\ref{2.1}), (\ref{2.3}), (\ref{2.11}) and (\ref{2.12}). It depends on the
tensorial coefficients $\bfsfF_1$, $\bfsfF_2$ and implicitly on $\bfsfT$
and $\bfsfS$ through ${\bf P}$ and ${\bf W}$. All of these coefficients can
as yet be chosen at will, as long as the identities
(\ref{2.8})--(\ref{2.10}) are satisfied. To derive constraints on these
coefficients we turn to the equations of motion.

The Heisenberg equations of motion that follow from the total Hamiltonian 
are
\begin{eqnarray}
\fl \dot{\bf A}({\bf r},t)=\frac{1}{\varepsilon_0}\, \bPi({\bf r},t)
\label{2.13}\\
\fl \dot{\bPi}({\bf r},t)=\frac{1}{\mu_0}\, \Delta {\bf A}({\bf r},t)
+\hbarit \int d{\bf r}' \, {\bf W}({\bf r}',t)\cdot
\left[\bfsfF_1({\bf r}',{\bf r})\right]_T
-\hbarit \int d{\bf r}' \, 
\left[ \bfsfF_2({\bf r},{\bf r}')\right]_T\cdot {\bf A}({\bf r}',t)
\nonumber\\
 \label{2.14}\\
\fl \dot{\bf C}_m({\bf r},\omega,t)=
-\rmi\, \omega \, {\bf C}_m({\bf r},\omega,t)
-\rmi\, \omega \int d{\bf r}'\int d{\bf r}'' \,
\bfsfS^\ast({\bf r},{\bf r}',\omega)\cdot\bfsfF_1({\bf r}',{\bf r}'')\cdot
{\bf A}({\bf r}'',t)\nonumber\\
+\frac{1}{\varepsilon_0}\int d{\bf r}' \, \bfsfT^\ast({\bf r},{\bf
  r}',\omega)
\cdot [{\bf P}({\bf r}',t)]_{L'}
\label{2.15}
\end{eqnarray}
where all operators now depend on time. The subscript $L'$ denotes the
longitudinal part with respect to ${\bf r}'$. The time derivative of the
polarization density follows by combining (\ref{2.5}) and (\ref{2.15}):
\begin{eqnarray}
\fl \dot{\bf P}({\bf r},t)=-\hbarit \int d{\bf r}' \int_0^\infty d\omega'\, 
\omega'\, {\bf C}_m({\bf r}',\omega',t)\cdot\bfsfT({\bf r}',{\bf r}, \omega')\nonumber\\
\fl -\hbarit\int d{\bf r}'\int d{\bf r}''\int d{\bf r}'''\int_0^\infty
d\omega'\, \omega' \, \tilde{\bfsfT}({\bf r}',{\bf
  r},\omega')\cdot\bfsfS^\ast({\bf r}',{\bf r}'',\omega')\cdot
\bfsfF_1({\bf r}'',{\bf r}''')\cdot{\bf A}({\bf r}''',t)\nonumber\\
\fl -\rmi\,\frac{\hbarit}{\varepsilon_0}\int d{\bf r}'\int d{\bf r}''
\int_0^\infty d\omega' \, 
\tilde{\bfsfT}({\bf r}',{\bf r},\omega')\cdot
\bfsfT^\ast({\bf r}',{\bf r}'',\omega')\cdot
[{\bf P}({\bf r}'',t)]_{L''} +{\rm h.c.}
\label{2.16}
\end{eqnarray}
where h.c.\ denotes the Hermitian conjugate of all preceding terms. Upon
using (\ref{2.9}) one finds that the second term (together with its
Hermitian conjugate) equals $-\hbarit\int d{\bf r}'\, \bfsfF_1({\bf r},{\bf
r}')\cdot{\bf A}({\bf r}',t)$. Furthermore, the last term (again together
with its Hermitian conjugate) vanishes on account of (\ref{2.8}).

Eliminating $\bPi$ from (\ref{2.13}) and (\ref{2.14}) we find an
inhomogeneous wave equation for the vector potential
\begin{eqnarray}
\Delta{\bf A}({\bf r},t)-\frac{1}{c^2}\ddot{\bf A}({\bf r},t)=
-\mu_0\, \hbarit \int d{\bf r}'\, 
{\bf W}({\bf r}',t)\cdot
\left[\bfsfF_1({\bf r}', {\bf r})\right]_T \nonumber\\
+\mu_0\, \hbarit \int d{\bf r}' \,\left[\bfsfF_2({\bf r},{\bf
    r}')\right]_T\cdot{\bf A}({\bf r}',t)
\label{2.17}
\end{eqnarray}
where the first term at the right-hand side can be be expressed in terms of
the medium operators by substituting (\ref{2.6}). According to the Maxwell
equations the vector-potential source term, which is given by the
right-hand side of (\ref{2.17}), should equal $-\mu_0 \, [\dot{\bf P}({\bf
r},t)]_T$. Hence, comparison with (\ref{2.16}) leads to the identity
\begin{eqnarray}
\fl -\hbarit \int d{\bf r}'\int d{\bf r}''\int_0^\infty d\omega'\, \omega'
\left\{ {\bf C}_m({\bf r}',\omega',t)\cdot\bfsfS({\bf r}',{\bf
    r}'',\omega')\right.\nonumber\\
\fl \left. +{\bf C}_m^\dagger({\bf r}',\omega',t)\cdot
\bfsfS^\ast({\bf r}',{\bf r}'',\omega')\right\}\cdot\left[\bfsfF_1({\bf
    r}'',{\bf r})\right]_T
-\hbarit\int d{\bf r}'\,\left[\bfsfF_2({\bf
    r},{\bf r}'\right]_T\cdot{\bf A}({\bf r}',t)=\nonumber\\
\fl =-\hbarit\int d{\bf r}'\int_0^\infty d\omega'\, \omega'\,
\left\{{\bf C}_m({\bf r}',\omega',t)\cdot\left[\bfsfT({\bf r}',{\bf r},\omega')\right]_T
\right.\nonumber\\
\left.+{\bf C}^\dagger_m({\bf r}',\omega',t)\cdot\left[\bfsfT^\ast({\bf r}',{\bf
  r},\omega')\right]_T\right\}
-\hbarit\int d{\bf r}'\, \left[\bfsfF_1({\bf r},{\bf r}')\right]_T
\cdot{\bf A}({\bf r}',t).
\label{2.18}
\end{eqnarray}
Upon equating the coefficient of the vector potential we arrive at the
relation $\left[\bfsfF_1({\bf r},{\bf r}')\right]_{TT'}=
\left[\bfsfF_2({\bf r},{\bf r}')\right]_{TT'}$, which connects the
transverse parts of the tensors $\bfsfF_1$ and $\bfsfF_2$. A second
relation, namely $[\bfsfT({\bf r},{\bf r}',\omega)]_{T'}=\int d{\bf r}''\, 
\bfsfS({\bf r},{\bf r}'',\omega)\cdot [\bfsfF_1({\bf r}'',{\bf r}')]_{T'}$, 
follows by equating the coefficient of ${\bf C}_m$. Combining
these two relations with (\ref{2.8})--(\ref{2.10}) we thus have found that
the tensors $\bfsfT$, $\bfsfS$, $\bfsfF_1$ and $\bfsfF_2$ have to satisfy
five conditions. Apart from these constraints the coefficients can be
chosen freely while constructing our model Hamiltonian. We shall use this
freedom to impose instead of the relations following from (\ref{2.18}) two
somewhat stronger conditions that result upon including the longitudinal
parts:
\begin{eqnarray}
\bfsfF_1({\bf r},{\bf r}')=\bfsfF_2({\bf r},{\bf r}')
\label{2.19}\\
\bfsfT({\bf r},{\bf r}',\omega)=\int d{\bf r}''\, 
\bfsfS({\bf r},{\bf r}'',\omega)\cdot \bfsfF_1({\bf r}'',{\bf r}').
\label{2.20}
\end{eqnarray}
As the first of these equalities shows, the bilinear coupling of ${\bf W}$
with ${\bf A}$ and the quadratic vector-potential term in (\ref{2.11}) have
to occur simultaneously. This is a well-known consequence of the
minimal-coupling scheme. In view of (\ref{2.19}) we shall omit the
subscripts of $\bfsfF_i$ in the following. Furthermore, we shall use
(\ref{2.20}) to eliminate $\bfsfS$ from the formalism altogether.

Summarizing the above results, we have obtained the following Hamiltonian
for a linear inhomogeneous anisotropic dispersive dielectric interacting
with the electromagnetic field:
\begin{eqnarray}
\fl H=\int d{\bf r} \left\{\frac{1}{2\varepsilon_0}
[\bPi({\bf r})]^2+\frac{1}{2\mu_0}[\bnabla\wedge{\bf A}({\bf r})]^2\right\}
+\hbarit\int d{\bf r}\int_0^\infty d\omega \; \omega \, 
{\bf  C}_m^{\dagger}({\bf r},\omega)\cdot {\bf  C}_m({\bf
  r},\omega)\nonumber\\
\fl + \,\hbarit\int d{\bf r}\int d{\bf r}'\int_0^\infty d\omega\, \omega \left[
{\bf C}_m({\bf r},\omega)\cdot\bfsfT({\bf r},{\bf r}',\omega)+
{\bf C}_m^\dagger({\bf r},\omega)\cdot\bfsfT^\ast({\bf r},{\bf
  r}',\omega)\right]
\cdot{\bf A}({\bf r}')\nonumber\\
+\half \, \hbarit \int d{\bf r}\int d{\bf r}' {\bf A}({\bf r})\cdot 
\bfsfF({\bf r},{\bf r}')\cdot{\bf A}({\bf r}')
+ \frac{1}{2\varepsilon_0}\int d{\bf r} \, \left\{[{\bf P}({\bf r})]_L\right\}^2.
\label{2.21}
\end{eqnarray}
The complex tensorial coefficient $\bfsfT$ can be chosen freely. It has to
satisfy two constraints, the first of which has been written already in
(\ref{2.8}). The second one follows by substituting (\ref{2.20}) in
(\ref{2.10}):
\begin{equation}
\int d{\bf r}''\int_0^\infty d\omega'' \, \omega''^2\, 
\tilde{\bfsfT}({\bf r}'',{\bf r},\omega'')\cdot
{\bfsfT}^\ast({\bf r}'',{\bf r}',\omega'')-{\rm c.c.}=0.
\label{2.22}
\end{equation}
Finally, insertion of (\ref{2.20}) in (\ref{2.9}) leads to the equality
\begin{equation}
 \int d{\bf r}''\int_0^\infty d\omega''\, \omega''\,
\tilde{\bfsfT}({\bf r}'',{\bf r}, \omega'')\cdot
\bfsfT^\ast({\bf r}'',{\bf r}',\omega'')+{\rm c.c.}=\bfsfF({\bf r},{\bf r}').
\label{2.23}
\end{equation}
This relation defines the real tensor $\bfsfF$ in terms of $\bfsfT$. It
shows that $\bfsfF({\bf r},{\bf r}')$ satisfies the symmetry property
$\tilde{\bfsfF}({\bf r},{\bf r}')= \bfsfF({\bf r}',{\bf r})$, as we know
already from the way $\bfsfF_2$ occurs in (\ref{2.11}). As an integral
kernel the tensor $\bfsfF({\bf r},{\bf r}')$ is positive-definite. This is
established by taking the scalar products of (\ref{2.23}) with real vectors
${\bf v}({\bf r})$ and ${\bf v}({\bf r}')$, and integrating over ${\bf r}$
and ${\bf r}'$. The result is positive for any choice of ${\bf v}$. As a
consequence, the inverse of $\bfsfF$ is well defined.

The polarization density is given by (\ref{2.5}), while the canonical momentum
density reads according to (\ref{2.6}) with (\ref{2.20}):
\begin{equation}
\fl {\bf W}({\bf r})=-\int d{\bf r}'\int d{\bf r}''\int_0^\infty
d\omega'\, \omega' \, 
{\bf C}_m({\bf r}',\omega')\cdot
\bfsfT({\bf r}',{\bf  r}'',\omega')\cdot 
\bfsfF^{-1}({\bf r}'',{\bf r})+{\rm h.c.}
\label{2.24}
\end{equation}
where the right-hand side contains the inverse of $\bfsfF$.

The Hamiltonian (\ref{2.21}) has been constructed by starting from general
forms for its parts $H_f$, $H_m$, $H_i$ and $H_{\rm es}$ and requiring
consistency with Maxwell's equations. It may be related to a Lagrange
formalism, as is shown in Appendix A.

In the following we shall investigate the dynamics of the model defined by
(\ref{2.21}). As the Hamiltonian is quadratic in the dynamical variables it
is possible to accomplish a complete diagonalization. This will be the
subject of the next section.

\section{Diagonalization of the Hamiltonian}\label{sec3}
We wish to find a diagonal representation of the Hamiltonian (\ref{2.21})
in the form
\begin{equation}
H=\hbarit\int d{\bf r}\int_0^\infty d\omega\, \omega \,
{\bf C}^{\dagger}({\bf r}, \omega)\cdot {\bf C}({\bf r}, \omega). 
\label{3.1}
\end{equation}
The creation and annihilation operators satisfy the standard commutation
relations of the form (\ref{2.4}). They are linear combinations of the
dynamical variables in (\ref{2.21}): 
\begin{eqnarray}
\fl {\bf C}({\bf r},\omega)=\int d{\bf r}'\left\{ \rule{0mm}{5mm}
\bfsff_1({\bf r},{\bf r}',\omega)\cdot{\bf A}({\bf r}')
+\bfsff_2({\bf r},{\bf r}',\omega)\cdot \bPi({\bf
  r}')\right.\nonumber\\
\fl +\,\left. \int_0^\infty d\omega'\left[
\bfsff_3({\bf r},{\bf r}',\omega,\omega')\cdot{\bf C}_m({\bf r}',\omega')
+\bfsff_4({\bf r},{\bf r}',\omega,\omega')\cdot
{\bf C}_m^{\dagger}({\bf r}',\omega')
\right]\right\}
\label{3.2}
\end{eqnarray}
with as-yet unknown tensorial coefficients $\bfsff_i$, the first two of which
are taken to be transverse in their second argument. To determine $\bfsff_i$
we use Fano's method \cite{F61}: we evaluate the commutator $[{\bf
C}({\bf r},\omega),H]$ and equate the result to $\hbarit\,\omega \, {\bf
C}({\bf r},\omega)$. Comparing the contributions involving the various
canonical operators we arrive at the four equations
\begin{eqnarray}
\fl \frac{\rmi}{\varepsilon_0}\, \bfsff_1({\bf r},{\bf r}',\omega)=\omega\, 
\bfsff_2({\bf r},{\bf r}',\omega)
\label{3.3}\\
\fl \frac{\rmi}{\mu_0}\, \Delta'\bfsff_2({\bf r},{\bf r}',\omega)
-\rmi\,\hbarit\int d{\bf r}''\, \bfsff_2({\bf r},{\bf r}'',\omega)
\cdot\left[\bfsfF({\bf r}'',{\bf r}')\right]_{T'}\nonumber\\
\fl +\, \int d{\bf r}''\int_0^\infty d\omega''\, \omega''\left\{
\bfsff_3({\bf r},{\bf r}'',\omega,\omega'')\cdot
[\bfsfT^\ast({\bf r}'',{\bf r}',\omega'')]_{T'}\right.\nonumber\\
 \left. -\bfsff_4({\bf r},{\bf r}'',\omega,\omega'')\cdot
[\bfsfT({\bf r}'',{\bf r}',\omega'')]_{T'}\right\}=\omega\, \bfsff_1({\bf
  r},{\bf r}',\omega)
\label{3.4}\\
\fl -\rmi\, \hbarit\,\omega'\int d{\bf r}''\, 
\bfsff_2({\bf r},{\bf r}'',\omega)\cdot
\tilde{\bfsfT}({\bf r}',{\bf  r}'',\omega')
+\omega'\, \bfsff_3({\bf r},{\bf r}',\omega,\omega')\nonumber\\
\fl +\frac{\hbarit}{\varepsilon_0}\int d{\bf r}''\int d{\bf r}'''
\int_0^\infty d\omega''
\left\{\bfsff_3({\bf r},{\bf r}'',\omega,\omega'')
\cdot[\bfsfT^\ast({\bf r}'',{\bf r}''',\omega'')]_{L'''}\right.\nonumber\\
\fl \left. +\bfsff_4({\bf r},{\bf r}'',\omega,\omega'')
\cdot[\bfsfT({\bf r}'',{\bf r}''',\omega'')]_{L'''}\right\}
\cdot\tilde{\bfsfT}({\bf r}',{\bf r}''',\omega')
=\omega\,\bfsff_3({\bf r},{\bf r}',\omega,\omega')
\label{3.5}\\
\fl -\rmi\, \hbarit\,\omega'\int d{\bf r}''\, 
\bfsff_2({\bf r},{\bf r}'',\omega)\cdot
\tilde{\bfsfT}^\ast({\bf r}',{\bf  r}'',\omega')
-\omega'\, \bfsff_4({\bf r},{\bf r}',\omega,\omega')\nonumber\\
\fl -\frac{\hbarit}{\varepsilon_0}\int d{\bf r}''\int d{\bf r}'''
\int_0^\infty d\omega''
\left\{\bfsff_3({\bf r},{\bf r}'',\omega,\omega'')
\cdot[\bfsfT^\ast({\bf r}'',{\bf r}''',\omega'')]_{L'''}\right.\nonumber\\
\fl \left. +\bfsff_4({\bf r},{\bf r}'',\omega,\omega'')
\cdot[\bfsfT({\bf r}'',{\bf r}''',\omega'')]_{L'''}\right\}
\cdot\tilde{\bfsfT}^\ast({\bf r}',{\bf r}''',\omega')
=\omega\,\bfsff_4({\bf r},{\bf r}',\omega,\omega').
\label{3.6}
\end{eqnarray}

The solution of these equations can be obtained by a method that is a
generalization of that used in our earlier work \cite{SWo04}. The details
are given in appendices B and C. The results are:
\begin{eqnarray}
\fl \bfsff_1({\bf r},{\bf r}',\omega)=\frac{\omega^2}{c^2}
\int d{\bf r}'' \, 
\bfsfT^\ast({\bf r},{\bf r}'',\omega)\cdot
[\bfsfG({\bf r}'',{\bf r}',\omega-\rmi\, 0)]_{T'}
\label{3.7}\\
\fl \bfsff_2({\bf r},{\bf r}',\omega)=\rmi\, \mu_0\, \omega
\int d{\bf r}'' \, 
\bfsfT^\ast({\bf r},{\bf r}'',\omega)\cdot
[\bfsfG({\bf r}'',{\bf r}',\omega-\rmi\, 0)]_{T'}
\label{3.8}\\
\fl \bfsff_3({\bf r},{\bf r}',\omega,\omega')=
\bfsfI \,  \delta({\bf r}-{\bf r}')\, \delta(\omega-\omega')
 -\mu_0\,\hbarit\,\omega \int d{\bf r}''\int d{\bf r}''' 
\, \bfsfT^\ast({\bf r},{\bf r}'',\omega) \nonumber\\
\cdot [\bfsfG({\bf r}'',{\bf r}''',\omega-\rmi \, 0)]_{T'''}\cdot
\tilde{\bfsfT}({\bf r}',{\bf r}''',\omega')\nonumber\\
 +\mu_0\,\hbarit\,\frac{\omega^2}{\omega-\omega'-\rmi\, 0}
\int d{\bf r}''\int d{\bf r}''' \, 
\bfsfT^\ast({\bf r},{\bf r}'',\omega)\nonumber\\
\cdot \bfsfG({\bf r}'',{\bf r}''',\omega-\rmi\, 0) \cdot
\tilde{\bfsfT}({\bf r}',{\bf r}''',\omega')
\label{3.9}\\
\fl \bfsff_4({\bf r},{\bf r}',\omega,\omega')=
\mu_0\,\hbarit\,\omega \int d{\bf r}''\int d{\bf r}''' 
\, \bfsfT^\ast({\bf r},{\bf r}'',\omega) 
\cdot [\bfsfG({\bf r}'',{\bf r}''',\omega-\rmi \, 0)]_{T'''}\cdot
\tilde{\bfsfT}^\ast({\bf r}',{\bf r}''',\omega')\nonumber\\
 -\mu_0\,\hbarit\,\frac{\omega^2}{\omega+\omega'}
\int d{\bf r}''\int d{\bf r}''' \, 
\bfsfT^\ast({\bf r},{\bf r}'',\omega)\nonumber\\
\cdot \bfsfG({\bf r}'',{\bf r}''',\omega-\rmi\, 0) \cdot
\tilde{\bfsfT}^\ast({\bf r}',{\bf r}''',\omega').
\label{3.10}
\end{eqnarray}

The Green function $\bfsfG({\bf r},{\bf r}',z)$ occurring in these
expressions is defined as the solution of the differential equation:
\begin{eqnarray}
 -\left[\bfsfG({\bf r},{\bf r}',z)\times \overleftarrow{\bnabla}'\right]
\times \overleftarrow{\bnabla}'
+\frac{z^2}{c^2}\, \bfsfG({\bf r},{\bf r}',z)\nonumber\\
 +\frac{z^2}{c^2}\int d{\bf r}''\, \bfsfG({\bf r},{\bf r}'',z)\cdot
\bchi({\bf r}'',{\bf r}',z)=\bfsfI\, \delta({\bf
  r}-{\bf r}')
\label{3.11}
\end{eqnarray}
The spatial derivative operator $\bnabla'$ acts to the left on the argument
${\bf r}'$ of $\bfsfG({\bf r},{\bf r}',z)$.  According to this
inhomogeneous wave equation the Green function determines the propagation
of waves through a medium that is characterized by a tensor $\bchi({\bf
r},{\bf r}',z)$. The latter plays the role of a non-local anisotropic
susceptibility, as will become clear in the next section. It is defined in
terms of $\bfsfT$ and its complex conjugate as
\begin{eqnarray}
\bchi({\bf r},{\bf r}',z)\equiv
\frac{\hbarit}{\varepsilon_0}\int d{\bf r}''\int_0^\infty d\omega
\left[\frac{1}{\omega-z}\,\tilde{\bfsfT}({\bf r}'',{\bf r},\omega)\cdot
\bfsfT^\ast({\bf r}'',{\bf r}',\omega)\right.\nonumber\\
\left.+\frac{1}{\omega+z}\,\tilde{\bfsfT}^\ast({\bf r}'',{\bf r},\omega)\cdot
\bfsfT({\bf r}'',{\bf r}',\omega)\right]
\label{3.12}
\end{eqnarray}
with the frequency argument $z$ either in the upper half or the lower half
of the complex $z$-plane, which has got a cut along the real axis.
Likewise, the Green function in (\ref{3.11}) is defined in the complex
$z$-plane with a cut along the real axis. Both the susceptibility and the
Green function are discontinuous across the cut.

We have succeeded now in finding the diagonal representation of the
Hamiltonian of our model. The diagonalizing operators are given by
(\ref{3.2}), with coefficients that are listed in
(\ref{3.7})--(\ref{3.10}). 

\section{Field, polarization and susceptibility}\label{sec4}
Once we have the diagonal representation of the Hamiltonian at our
disposal, we can determine the full time evolution of the dynamical
variables. In the following we will derive the time dependence of the
vector potential, the electric field and the polarization. As we shall need
a few properties of the tensors $\bchi$ and $\bfsfG$, we shall
discuss these first.

From its definition (\ref{3.12}) it follows that the tensor $\bchi$
satisfies the symmetry relations
\begin{equation}
\tilde{\bchi}({\bf r},{\bf r}',z)=
\bchi({\bf r}',{\bf r},-z)
\label{4.1}
\end{equation}
and
\begin{equation}
\bchi^\ast({\bf r},{\bf r}',z)=
\bchi({\bf r},{\bf r}',-z^\ast)
\label{4.2}
\end{equation}
so that $\bchi({\bf r},{\bf r}',z)$ is real on the
imaginary axis. The discontinuity across the cut along the real axis is
given by
\begin{equation}
\fl\bchi({\bf r},{\bf r}',\omega+\rmi\, 0)-
\bchi({\bf r},{\bf r}',\omega-\rmi\, 0)=
\frac{2\,\pi\,\rmi\,\hbarit}{\varepsilon_0}
\int d{\bf r}''\, \tilde{\bfsfT}({\bf r}'',{\bf r},\omega)\cdot
\bfsfT^\ast({\bf r}'',{\bf r}',\omega)
\label{4.3}
\end{equation}
for positive $\omega$ and by
\begin{equation}
\fl\bchi({\bf r},{\bf r}',\omega+\rmi\, 0)-
\bchi({\bf r},{\bf r}',\omega-\rmi\, 0)=
-\frac{2\,\pi\,\rmi\,\hbarit}{\varepsilon_0}
\int d{\bf r}''\, \tilde{\bfsfT}^\ast({\bf r}'',{\bf r},-\omega)\cdot
\bfsfT({\bf r}'',{\bf r}',-\omega)
\label{4.4}
\end{equation}
for negative $\omega$. Hence, we may write (\ref{3.12}) as
\begin{equation}
\fl\bchi({\bf r},{\bf r}',z)=
\frac{1}{2\,\pi\,\rmi}\int_{-\infty}^\infty d\omega\,
\frac{1}{\omega-z}\, \left[
\bchi({\bf r},{\bf r}',\omega+\rmi\, 0)-
\bchi({\bf r},{\bf r}',\omega-\rmi\, 0)\right]
\label{4.5}
\end{equation}
which is the well-known Kramers-Kronig relation for the Fourier transform
of a causal function. The identities (\ref{2.8}), (\ref{2.23}) and
(\ref{2.22}) can be rewritten in terms of the discontinuity across the cut:
\begin{eqnarray}
\int_{-\infty}^\infty d\omega\, 
\left[
\bchi({\bf r},{\bf r}',\omega+\rmi\, 0)-
\bchi({\bf r},{\bf r}',\omega-\rmi\, 0)\right]=0
\label{4.6}\\
\int_{-\infty}^\infty d\omega\, \omega\,
\left[
\bchi({\bf r},{\bf r}',\omega+\rmi\, 0)-
\bchi({\bf r},{\bf r}',\omega-\rmi\, 0)\right]=
\frac{2\,\pi\,\rmi\,\hbarit}{\varepsilon_0}\,\bfsfF({\bf r},{\bf r}')
\label{4.7}\\
\int_{-\infty}^\infty d\omega\, \omega^2\, 
\left[
\bchi({\bf r},{\bf r}',\omega+\rmi\, 0)-
\bchi({\bf r},{\bf r}',\omega-\rmi\, 0)\right]=0.
\label{4.8}
\end{eqnarray}
Incidentally, we remark that for large $|z|$ the asymptotic behaviour of
$\bchi$ follows from (\ref{4.5}) with (\ref{4.6})--(\ref{4.8}) as
\begin{equation}
\bchi({\bf r},{\bf r}',z)\simeq
-\frac{\hbarit}{\varepsilon_0}\,\bfsfF({\bf r},{\bf r}')\, \frac{1}{z^2}+
{\cal O}\left(\frac{1}{z^4}\right).
\label{4.9}
\end{equation}

The above symmetry properties can be used to prove analogous symmetry
relations for the Green function $\bfsfG$. By taking the complex conjugate
of (\ref{3.11}) and using (\ref{4.2}) one derives
\begin{equation}
\bfsfG^\ast({\bf r},{\bf r}',z)=\bfsfG({\bf r},{\bf r}',-z^\ast).
\label{4.10}
\end{equation}
The adjoint equation of (\ref{3.11}) reads
\begin{eqnarray}
 -\bnabla\times\left[\bnabla\times\bfsfG({\bf r},{\bf r}',z)\right]
+\frac{z^2}{c^2}\, \bfsfG({\bf r},{\bf r}',z)\nonumber\\
 +\frac{z^2}{c^2}\int d{\bf r}''\, 
\bchi({\bf r},{\bf r}'',z)\cdot
\bfsfG({\bf r}'',{\bf r}',z)
=\bfsfI\, \delta({\bf  r}-{\bf r}')
\label{4.11}
\end{eqnarray}
as follows from (\ref{3.11}) after multiplication by 
$\bfsfG({\bf r}',{\bf r}'',z)$, integration over ${\bf r}'$ and a partial
integration. Comparing this differential equation to that obtained by
taking the transpose of (\ref{3.11}) and interchanging the position
variables one finds with the use of (\ref{4.1}) the reciprocity relation:
\begin{equation}
\tilde{\bfsfG}({\bf r},{\bf r}',z)= 
\bfsfG({\bf r}',{\bf r},-z).
\label{4.12}
\end{equation}

Having obtained the relevant physical properties of the tensors $\bchi$ and
$\bfsfG$, we return to a discussion of the time dependence of the dynamical
variables. Inverting (\ref{3.2}) by means of the canonical commutation
relations we get:
\begin{equation}
{\bf A}({\bf r})=\rmi\, \hbarit\int d{\bf r}'\int_0^\infty d\omega\, \,
\tilde{\bfsff}^\ast_2({\bf r}',{\bf r},\omega)\cdot
{\bf C}({\bf r}',\omega) + {\rm h.c.}
\label{4.13}
\end{equation}
Substitution of (\ref{3.8}) yields with the help of (\ref{4.10}) and (\ref{4.12}):
\begin{eqnarray}
\fl {\bf A}({\bf r},t)= \mu_0\, \hbarit \int d{\bf r}'\int d{\bf r}''
\int_0^\infty d\omega\, \omega 
\left[\bfsfG({\bf r},{\bf r}',\omega+\rmi\, 0)\right]_T\nonumber\\
\cdot\tilde{\bfsfT}({\bf r}'',{\bf r}',\omega)
\cdot {\bf C}({\bf r}'',\omega)\, \rme^{-\rmi\,\omega \,t}+{\rm h.c.}
\label{4.14}
\end{eqnarray}
where we accounted for the time dependence of the diagonalizing
operator ${\bf C}({\bf r}'',\omega)$. 

By differentiation with respect to space and time we can easily determine
the time evolution of the electromagnetic fields . Taking the curl of
(\ref{4.14}) we get:
\begin{eqnarray}
\fl {\bf B}({\bf r},t)=\mu_0\, \hbarit \int d{\bf r}'\int d{\bf r}''
\int_0^\infty d\omega\, \omega \,
\bnabla\times\bfsfG({\bf r},{\bf r}',\omega+\rmi\, 0)\nonumber\\
\cdot\tilde{\bfsfT}({\bf r}'',{\bf r}',\omega)
\cdot {\bf C}({\bf r}'',\omega)\, \rme^{-\rmi\,\omega \,t}+{\rm h.c.}
\label{4.15}
\end{eqnarray}
Furthermore, the transverse part of the electric field follows from
(\ref{4.14}) by differentiation with respect to $t$:
\begin{eqnarray}
\fl \left[{\bf E}({\bf r},t)\right]_T= \rmi\,\mu_0\, \hbarit 
\int d{\bf r}'\int d{\bf r}''\int_0^\infty d\omega\, \omega^2 \,
\left[\bfsfG({\bf r},{\bf r}',\omega+\rmi\, 0)\right]_T\nonumber\\
\cdot\tilde{\bfsfT}({\bf r}'',{\bf r}',\omega)
\cdot {\bf C}({\bf r}'',\omega)\, \rme^{-\rmi\,\omega \,t}+{\rm h.c.}
\label{4.16}
\end{eqnarray}
To determine the longitudinal part of the electric field we first have to
derive an expression for the polarization. 

The medium operator ${\bf C}_m({\bf r},\omega)$ is a linear combination of
the diagonalizing operator and its Hermitian conjugate: 
\begin{equation}
\fl {\bf C}_m({\bf r},\omega)=\int d{\bf r}'\int_0^\infty d\omega'
\left[\tilde{\bfsff}^\ast_3({\bf r}',{\bf r},\omega',\omega)\cdot
{\bf C}({\bf r}',\omega')-
\tilde{\bfsff}_4({\bf r}',{\bf r},\omega',\omega)\cdot
{\bf C}^\dagger({\bf r}',\omega')\right]
\label{4.17}
\end{equation}
as follows by taking the inverse of (\ref{3.2}). Substituting
(\ref{3.9})--(\ref{3.10}) and inserting the result in (\ref{2.5}) we get
after some algebra
\begin{eqnarray}
\fl {\bf P}({\bf r},t)=\frac{\rmi\,\hbarit}{c^2}
\int d{\bf r}'\int d{\bf r}''\int d{\bf r}'''
\int_0^\infty d\omega\, \omega^2\, 
\bchi({\bf r},{\bf r}',\omega+\rmi\,0)
\cdot \bfsfG({\bf r}',{\bf r}'',\omega+\rmi\,0)\nonumber\\
\cdot \tilde{\bfsfT}({\bf r}''',{\bf r}'',\omega)\cdot {\bf C}({\bf r}''',\omega)\,
\rme^{-\rmi\,\omega\,t} \nonumber\\
-\rmi\,\hbarit\int d{\bf r}'\int_0^\infty d\omega\,
\tilde{\bfsfT}({\bf r}',{\bf r},\omega)\cdot{\bf C}({\bf r}',\omega)
\, \rme^{-\rmi\,\omega\, t} +{\rm h.c.}
\label{4.18}
\end{eqnarray}
where we have employed (\ref{2.8}) and (\ref{3.12}). The longitudinal part
of this expression can be rewritten with the use of (\ref{4.11}) as:
\begin{eqnarray}
\fl \left[{\bf P}({\bf r},t)\right]_L= -\frac{\rmi\,\hbarit}{c^2} \int
d{\bf r}'\int d{\bf r}'' \int_0^\infty d\omega\, \omega^2\, \left[
  \bfsfG({\bf r},{\bf r}',\omega+\rmi\,0)\right]_L \nonumber\\
\cdot\tilde{\bfsfT}({\bf r}'',{\bf r}',\omega)\cdot {\bf C}({\bf
  r}'',\omega)\, \rme^{-\rmi\,\omega\,t}+{\rm h.c.}
\label{4.19}
\end{eqnarray}
From the Maxwell equation $\bnabla\cdot(\varepsilon_0\,{\bf E}+{\bf P})=0$
it follows that the left-hand side is proportional to the longitudinal part
$[{\bf E}({\bf r},t)]_L$ of the electric field. The ensuing expression for
the latter is analogous to (\ref{4.16}), so that we arrive at the following
result for the complete electric field:
\begin{eqnarray}
\fl {\bf E}({\bf r},t)= \rmi\,\mu_0\, \hbarit 
\int d{\bf r}'\int d{\bf r}''\int_0^\infty d\omega\, \omega^2 \,
\bfsfG({\bf r},{\bf r}',\omega+\rmi\, 0)
\cdot\tilde{\bfsfT}({\bf r}'',{\bf r}',\omega)
\cdot {\bf C}({\bf r}'',\omega)\, \rme^{-\rmi\,\omega \,t}\nonumber\\
+{\rm h.c.}
\label{4.20}
\end{eqnarray}

Inspection of (\ref{4.18}) shows that the polarization consists of two
terms. The first term is proportional to the electric field, at least in
Fourier space and after taking a spatial convolution integral. The
proportionality factor is $\bchi({\bf r},{\bf r}',\omega)$, which plays the
role of a susceptibility tensor, as we anticipated in the previous
section. The second term in (\ref{4.18}) is not related to the electric
field. It represents a noise polarization density ${\bf P}_n({\bf r},t)$
defined as
\begin{equation}
{\bf P}_n({\bf r},t)=-\rmi\,\hbarit\int d{\bf r}'\int_0^\infty d\omega\,
\tilde{\bfsfT}({\bf r}',{\bf r},\omega)\cdot{\bf C}({\bf r}',\omega)
\, \rme^{-\rmi\,\omega\, t} +{\rm h.c.}
\label{4.21}
\end{equation}
that has to be present so as to yield a quantization scheme in which the
validity of the canonical commutation relations in the presence of
dissipation is guaranteed. Introducing the Fourier transform ${\bf
P}_n({\bf r},\omega)$ via
\begin{equation}
{\bf P}_n({\bf r},t)=\int_0^\infty d\omega\, 
{\bf P}_n({\bf r},\omega)\,
\rme^{-\rmi\, \omega\, t} + {\rm h.c.}
\label{4.22}
\end{equation}
and its counterparts ${\bf E}({\bf r},\omega)$ and ${\bf P}({\bf r},\omega)$,
we get from (\ref{4.18}) with (\ref{4.20}):
\begin{equation}
{\bf P}({\bf r},\omega)=\int d{\bf r}'
\bchi({\bf r},{\bf r}',\omega+\rmi\,0)\cdot{\bf E}({\bf r}',\omega)+
{\bf P}_n({\bf r},\omega).
\label{4.23}
\end{equation}

The Fourier-transformed noise polarization density is proportional to the
diagonalizing operator:
\begin{equation}
{\bf P}_n({\bf r},\omega)=-\rmi\,\hbarit\int d{\bf r}'\,
\tilde{\bfsfT}({\bf r}',{\bf r},\omega)\cdot{\bf C}({\bf r}',\omega).
\label{4.24}
\end{equation}
as follows from (\ref{4.21}) and (\ref{4.22}). As we have got now an
explicit expression for ${\bf P}_n({\bf r},\omega)$ we can derive its
commutation relation. By employing (\ref{4.3}) we obtain:
\begin{equation}
\fl \left[ {\bf P}_n({\bf r},\omega), {\bf P}^\dagger_n({\bf r}',\omega')\right]=
-\frac{\rmi\,\hbarit\,\varepsilon_0}{2\,\pi}\, 
\left[
\bchi({\bf r},{\bf r}',\omega+\rmi\, 0)-
\bchi({\bf r},{\bf r}',\omega-\rmi\, 0)\right]\,
\delta(\omega-\omega').
\label{4.25}
\end{equation}
This commutation relation is a generalization of that postulated in the
phenomenological quantization scheme for isotropic dielectrics without
spatial dispersion \cite{GW95}--\cite{KSW01}. In the present approach, 
we have been able to prove its validity.

Both the fields and the polarization density can be rewritten in terms of
${\bf P}_n({\bf r},\omega)$. We get from (\ref{4.15}), (\ref{4.18}) and
(\ref{4.20}) upon eliminating ${\bf C}({\bf r},\omega)$ in favour of ${\bf
  P}_n({\bf r},\omega)$: 
\begin{eqnarray}
\fl {\bf E}({\bf r},t)= -\mu_0
\int d{\bf r}'\int_0^\infty d\omega\, \omega^2 \,
\bfsfG({\bf r},{\bf r}',\omega+\rmi\, 0)
\cdot {\bf P}_n({\bf r}',\omega)\, \rme^{-\rmi\,\omega \,t}+{\rm h.c.}
\label{4.26}\\
\fl {\bf B}({\bf r},t)= \rmi\,\mu_0
\int d{\bf r}'\int_0^\infty d\omega\, \omega \,
\bnabla\times\bfsfG({\bf r},{\bf r}',\omega+\rmi\, 0)
\cdot {\bf P}_n({\bf r}',\omega)\, \rme^{-\rmi\,\omega \,t}+{\rm h.c.}
\label{4.27}\\
\fl {\bf P}({\bf r},t)=-\frac{1}{c^2}
\int d{\bf r}'\int d{\bf r}''
\int_0^\infty d\omega\, \omega^2\, 
\bchi({\bf r},{\bf r}',\omega+\rmi\,0)\cdot 
\bfsfG({\bf r}',{\bf r}'',\omega+\rmi\,0)
\cdot {\bf P}_n({\bf r}'',\omega)\,\rme^{-\rmi\,\omega\,t} \nonumber\\
+\int_0^\infty d\omega\, 
{\bf P}_n({\bf r},\omega)\,
\rme^{-\rmi\, \omega\, t} + {\rm h.c.}
\label{4.28}
\end{eqnarray}
By adding (\ref{4.26}) and (\ref{4.28}) we get an expression for the
dielectric displacement ${\bf D}({\bf r},t)$. Upon using (\ref{4.11}) we may
write it as
\begin{equation}
\fl {\bf D}({\bf r},t)= 
-\int d{\bf r}'\int_0^\infty d\omega\, 
\bnabla\times\left[\bnabla\times\bfsfG({\bf r},{\bf r}',\omega+\rmi\, 0)\right]
\cdot {\bf P}_n({\bf r}',\omega)\, \rme^{-\rmi\,\omega \,t}+{\rm h.c.}
\label{4.29}
\end{equation}
Clearly, the dielectric displacement is purely transverse. Comparing with
(\ref{4.27}) we find that Maxwell's equation $\bnabla\times{\bf B}({\bf
r},t)=\mu_0\, \partial{\bf D}({\bf r},t)/\partial t$ is satisfied.

It is instructive to return to the time-dependent representation of the
linear constitutive relation (\ref{4.23}):
\begin{equation}
{\bf P}({\bf r},t)=\int d{\bf r}'\int^t_{-\infty}dt'\, 
\bchi({\bf r},{\bf r}',t-t')\cdot{\bf E}({\bf r}',t')+
{\bf P}_n({\bf r},t) 
\label{4.30}
\end{equation}
with the time-dependent susceptibility tensor defined by writing:
\begin{equation}
\bchi({\bf r},{\bf r}',\omega+\rmi\, 0)=
\int_0^\infty dt\, \bchi({\bf r},{\bf r}',t)\, \rme^{\rmi\,\omega\,t}.
\label{4.31}
\end{equation}
The convolution integral in the first term of (\ref{4.30}), which expresses
the causal response of the medium, depends on the electric field at all
times $t'$ preceding $t$ and at all positions ${\bf r'}$, whereas the
second contribution is the noise term, which in classical theory does not
appear. Sometimes \cite{KA04} a different splitting of the various
contributions to the polarization density is proposed, by writing an
equation of the general form of (\ref{4.30}) in which the response term
covers only a limited range of values of $t'$, for instance $t'\in[0,t]$
for $t>0$. In such a formulation the convolution integral does not
represent the full causal response of the medium, so that part of the
response is hidden in the second term. As a consequence, the latter is no
longer a pure noise term, so that it cannot be omitted in the classical
version of the theory.

The above expressions for the fields and the polarization density in terms
of the Fourier-transformed noise polarization density satisfying the
commutation relations (\ref{4.25}) are the central results in the present
formalism for field quantization in inhomogeneous anisotropic
dielectric media with spatio-temporal dispersion. Although we are
describing dissipative media, it has not been necessary to explicitly
introduce a bath, as is commonly done in the context of damped-polariton
treatments \cite{HB92a,HB92b,SWo04,SWu04}. In the next section, we shall
show how a bath may be identified in the present model. 

\section{Bath degrees of freedom}\label{sec5}
In the Hamiltonian (\ref{2.21}) the dielectric medium is described by the
operators ${\bf C}_m({\bf r},\omega)$ and ${\bf C}^\dagger_m({\bf
r},\omega)$. The polarization density ${\bf P}({\bf r})$ and its canonical
conjugate ${\bf W}({\bf r})$ are given in (\ref{2.5}) and (\ref{2.24}) as
suitable linear combinations of the medium operators ${\bf C}_m$ and ${\bf
C}^\dagger_m$. Since the latter depend on the continuous variable $\omega$,
they describe many more degrees of freedom than ${\bf P}$ and ${\bf
W}$. The extra degrees of freedom can be taken together to define a
so-called `bath', which is independent of ${\bf P}$ and ${\bf W}$. Although
the name might suggest otherwise, the bath as introduced in this way is
part of the medium itself, and not some external environment. Its role is
to account for the dissipative effects in the dispersive medium, which may
arise for instance through a leak of energy by heat production. In the
following we shall identify the operators associated to the
bath. Subsequently, we shall show how the Hamiltonian can be rewritten so
as to give an explicit description of the coupling between the polarization
and the bath. In this way, we will be able to compare our model to its
counterparts in previous papers \cite{HB92a,HB92b,SWo04,SWu04}.

The bath will be described by operators ${\bf C}_b({\bf r},\omega)$ and 
${\bf C}^\dagger_b({\bf r},\omega)$ satisfying the usual commutation
relations. These bath operators are linear combinations of the medium
operators:
\begin{eqnarray}
\fl {\bf C}_b({\bf r},\omega)=\int d{\bf r}'\int_0^\infty d\omega' \,
\left[ 
\bfsfH_1({\bf r},{\bf r}',\omega,\omega')\cdot {\bf C}_m({\bf r}',\omega')
+\bfsfH_2({\bf r},{\bf r}',\omega,\omega')\cdot 
{\bf C}^\dagger_m({\bf r}',\omega')\right]\nonumber\\
\label{5.1}
\end{eqnarray}
with tensor coefficients $\bfsfH_i$ that will be determined
presently. Since the bath variables are by definition independent of both
${\bf P}({\bf r}')$ and ${\bf W}({\bf r}')$ for all ${\bf r}'$, they have to
commute with the latter. With the use of (\ref{2.5}) and (\ref{2.24}) we
get from these commutation relations the following conditions:
\begin{eqnarray}
\fl \int d{\bf r}''\int_0^\infty d\omega''\left[
\bfsfH_1({\bf r},{\bf r}'',\omega,\omega'')\cdot
\bfsfT^\ast({\bf r}'',{\bf r}',\omega'')
+\bfsfH_2({\bf r},{\bf r}'',\omega,\omega'')\cdot
\bfsfT({\bf r}'',{\bf r}',\omega'')\right]=0
\nonumber\\
\label{5.2}\\
\fl \int d{\bf r}''\int_0^\infty\omega''\, \omega''\left[
\bfsfH_1({\bf r},{\bf r}'',\omega,\omega'')\cdot
\bfsfT^\ast({\bf r}'',{\bf r}',\omega'')
-\bfsfH_2({\bf r},{\bf r}'',\omega,\omega'')\cdot
\bfsfT({\bf r}'',{\bf r}',\omega'')\right]=0.\nonumber\\
\label{5.3}
\end{eqnarray}

To determine $\bfsfH_i$ we start from the following {\em Ansatz}:
\begin{eqnarray}
\fl \bfsfH_1({\bf r},{\bf r}',\omega,\omega')=
\int d{\bf r}''\left[
\delta(\omega-\omega')\, \bfsfh_1({\bf r},{\bf r}'',\omega)+
\frac{1}{\omega-\omega'+\rmi\,0}\,\bfsfh_2({\bf r},{\bf r}'',\omega)
\right]\cdot\tilde{\bfsfT}({\bf r}',{\bf r}'',\omega')\nonumber\\
\label{5.4}\\
\fl \bfsfH_2({\bf r},{\bf r}',\omega,\omega')=
-\int d{\bf r}''\, \frac{1}{\omega+\omega'}\,
\bfsfh_2({\bf r},{\bf r}'',\omega)\cdot
\tilde{\bfsfT}^\ast({\bf r}',{\bf r}'',\omega')
\label{5.5}
\end{eqnarray}
with new tensor coefficients $\bfsfh_i$. Substituting these expressions in
(\ref{5.2})--(\ref{5.3}) and using (\ref{3.12}) and (\ref{4.3}), we find
that both of these conditions are simultaneously satisfied when $\bfsfh_1$
and $\bfsfh_2$ are related as
\begin{eqnarray}
\int d{\bf r}''\, \bfsfh_2({\bf r},{\bf r}'',\omega)\cdot
\bchi({\bf r}'',{\bf r}',\omega+\rmi\,0)=\nonumber\\
=\frac{1}{2\,\pi\,\rmi}\int d{\bf r}'' \,
\bfsfh_1({\bf r},{\bf r}'',\omega)\cdot
\left[ \bchi({\bf r}'',{\bf r}',\omega+\rmi\,0)-
\bchi({\bf r}'',{\bf r}',\omega-\rmi\,0)\right]. 
\label{5.6}
\end{eqnarray}
Hence, we are left with a single independent coefficient. It can be
determined by imposing the standard commutation relation of the form
(\ref{2.4}) for ${\bf C}_b({\bf r},\omega)$ and ${\bf C}_b^\dagger({\bf
r}',\omega')$. Using (\ref{5.1}) with (\ref{5.4})--(\ref{5.5}) we arrive at
the condition:
\begin{eqnarray}
\fl \frac{1}{2\,\rmi}\int d{\bf r}'' \int d{\bf r}'''\,
\bfsfh_1({\bf r},{\bf r}'',\omega)\cdot
\left[\bchi({\bf r}'',{\bf r}''',\omega+\rmi\,0)-
\bchi({\bf r}'',{\bf r}''',\omega-\rmi\,0)\right]\cdot
\tilde{\bfsfh}_1^\ast({\bf r}',{\bf r}''',\omega)=\nonumber\\
=\frac{\pi\,\hbarit}{\varepsilon_0}\,\bfsfI\, \delta({\bf r}-{\bf r}').
\label{5.7}
\end{eqnarray}
The vanishing of the commutator of ${\bf C}_b({\bf r},\omega)$ with ${\bf
C}_b({\bf r}',\omega')$ is warranted on the strength of (\ref{5.6}). 

In view of (\ref{4.3}) a solution of (\ref{5.7}) is 
\begin{equation}
\bfsfh_1({\bf r},{\bf r}',\omega)=\tilde{\bfsfT}^{-1}({\bf r}',{\bf
  r},\omega)
\label{5.8}
\end{equation}
and hence, on account of (\ref{5.6}):
\begin{equation}
\bfsfh_2({\bf r},{\bf r}',\omega)=
\frac{\hbarit}{\varepsilon_0}\int d{\bf r}''\,
\bfsfT^\ast({\bf r},{\bf  r}'',\omega)\cdot
\bchi^{-1}({\bf r}'',{\bf r}',\omega+\rmi\,0).
\label{5.9}
\end{equation}
It should be noted that the coefficients $\bfsfh_i$ are determined up to a
unitary transformation. This freedom, which is available to $\bfsfH_i$ as
well, corresponds to a natural arbitrariness in the choice of the bath operators
themselves. 

As the bath operators have been identified now, we can rewrite the
Hamiltonian so as to clarify their role in the dynamics of our model. To
that end we have to eliminate the medium operators ${\bf C}_m$ in favor of
the bath operators ${\bf C}_b$. Employing (\ref{2.5}), (\ref{2.24}) and
(\ref{5.1}) we can write the medium operators as:
\begin{eqnarray}
\fl {\bf C}_m({\bf r},\omega)=
\int d{\bf r}'\, \bfsfT^\ast({\bf r},{\bf r}',\omega) \cdot
\left[ \frac{\rmi}{\hbarit}\, \omega \int d{\bf r}''\, 
\bfsfF^{-1}({\bf r}',{\bf r}'')
\cdot{\bf P}({\bf r}'')-{\bf W}({\bf r}')\right]\nonumber\\
\fl +\int d{\bf r}'\int_0^\infty d\omega'\left[
\tilde{\bfsfH}_1^\ast({\bf r}',{\bf r},\omega',\omega)\cdot
{\bf C}_b({\bf r}',\omega')
-\tilde{\bfsfH}_2({\bf r}',{\bf r},\omega',\omega)
\cdot {\bf C}_b^\dagger({\bf r}',\omega')\right].
\label{5.10}
\end{eqnarray}
With the use of this expression the contributions involving the medium
operators in (\ref{2.21}) can be rewritten. In this way, we arrive at the 
following alternative form for the Hamiltonian of our model:
\begin{eqnarray}
\fl H=\int d{\bf r} \left\{\frac{1}{2\varepsilon_0}
[\bPi({\bf r})]^2+\frac{1}{2\mu_0}[\bnabla\wedge{\bf A}({\bf r})]^2\right\}
+\hbarit\int d{\bf r}\int_0^\infty d\omega \; \omega \, 
{\bf  C}_b^{\dagger}({\bf r},\omega)\cdot {\bf  C}_b({\bf
  r},\omega)\nonumber\\
\fl +\frac{\varepsilon_0}{2\,\pi\,\rmi\,\hbarit^2}
\int d{\bf r}\int d{\bf r}'\int d{\bf r}''\int d{\bf r}'''\,
{\bf P}({\bf r})\cdot\bfsfF^{-1}({\bf r},{\bf r}')
\nonumber\\
\fl\cdot\left\{\int_0^\infty 
d\omega\, \omega^3\left[ \bchi({\bf r}',{\bf r}'',\omega+\rmi\,0)-
\bchi({\bf r}',{\bf r}'',\omega-\rmi\,0)\right]\right\}\cdot
\bfsfF^{-1}({\bf r}'',{\bf r}''')\cdot{\bf P}({\bf r}''')
\nonumber\\
\fl + \frac{1}{2\varepsilon_0}\int d{\bf r} \, \left\{[{\bf P}({\bf
    r})]_L\right\}^2
+\half\, \hbarit \int d{\bf r}\int d{\bf r}' \,{\bf W}({\bf r})\cdot 
\bfsfF({\bf r},{\bf r}')\cdot{\bf W}({\bf r}')
\nonumber\\
\fl -\hbarit\int d{\bf r}\int d{\bf r}'\, {\bf W}({\bf r})\cdot
\bfsfF({\bf r},{\bf r}')\cdot{\bf A}({\bf r}')
+\half \, \hbarit \int d{\bf r}\int d{\bf r}' {\bf A}({\bf r})\cdot 
\bfsfF({\bf r},{\bf r}')\cdot{\bf A}({\bf r}')
\nonumber\\
\fl -\frac{\rmi\,\hbarit}{\varepsilon_0} 
\int d{\bf r} \int d{\bf r}' \int d{\bf r}'' \int_0^\infty d\omega \,
\left[
{\bf C}_b^\dagger({\bf r},\omega)\cdot
\bfsfT^\ast({\bf r},{\bf r}',\omega)\cdot
\bchi^{-1}({\bf r}',{\bf r}'',\omega+\rmi\, 0)\cdot{\bf P}({\bf r}'')
\right.\nonumber\\
\left.-{\rm h.c.}\right].
\label{5.11}
\end{eqnarray}
As before, the tensor coefficients $\bfsfT$ and $\bfsfF$ 
are related by (\ref{2.23}), while the susceptibility $\bchi$ follows from
$\bfsfT$ via (\ref{3.12}), which implies (\ref{4.3}).

We are now in a position to compare the present model with that discussed
in previous papers \cite{HB92a,HB92b,SWo04,SWu04}. The Hamiltonian
(\ref{5.11}) takes account of anisotropy and spatial dispersion. To make
contact with the earlier treatments these features should be left out. In
those circumstances both the susceptibility $\bchi$ and the tensor
coefficients $\bfsfT$, $\bfsfF$ are isotropic and local, so that one has
for instance:
\begin{equation}
\bchi({\bf r}, {\bf r}',z)=\chi({\bf r},z)\, \bfsfI\, 
\delta({\bf r}-{\bf r}').
\label{5.12}
\end{equation}
Furthermore, the dielectric medium of the present model has got an
arbitrary temporal dispersion: the frequency dependence of the
susceptibility is governed by that of $\bfsfT$, as is obvious from
(\ref{3.12}). In our previous treatment of the inhomogeneous
damped-polariton model \cite{SWo04,SWu04} the scalar susceptibility
satisfied a sum rule to the effect that the integral $\int_0^\infty
d\omega\, \omega^3\, [\chi({\bf r},\omega+\rmi\,0)- \chi({\bf
r},\omega-\rmi\, 0)]$, a generalization of which occurs in the third term
of (\ref{5.11}), is proportional to the square of an effective frequency
$\tilde{\omega}_0({\bf r})$. The latter parameter already figured in the
original Hamiltonian in \cite{HB92a,HB92b}, albeit in a space-independent
form. Finally, in our earlier work we followed the notation in
\cite{HB92a,HB92b} by representing the bath operators ${\bf C}_b({\bf
r},\omega)$ and their Hermitian conjugates by equivalent position and
momentum operators ${\bf Y}_\omega({\bf r})$ and ${\bf Q}_\omega({\bf
r})$. Implementing this alternative notation here as well, one shows that
(\ref{5.11}) indeed reduces to the Hamiltonian in \cite{SWo04,SWu04} for
the special case of an isotropic spatially-nondispersive medium.

\section{Conclusion}\label{sec6}

The Hamiltonian model that we have considered in this paper is a suitable
tool to underpin the quantum formalism for a general linear dielectric
medium and of the electromagnetic field propagating through such a
medium. By solving our model we have succeeded in giving a justification of
the postulates on which the phenomenological quantization scheme for
electrodynamics in dielectric media is usually based.

Our model incorporates many features to warrant the generality of the
description. Apart from allowing for inhomogeneities and anisotropies of
the medium it has the virtue of accommodating a quite general
spatio-temporal dispersion. In fact, the susceptibility tensor of the
dielectric medium has been identified in (\ref{3.12}), which implies the
Kramers-Kronig relation (\ref{4.5}). According to that relation the
susceptibility in the complex frequency plane is determined by the
discontinuity across the cut along the real axis. All anisotropic
inhomogeneous linear media with spatio-temporal dispersion that respond
causally to an external electric field are characterized by a
susceptibility tensor $\bchi({\bf r},{\bf r}',\omega)$ of the form
(\ref{4.5}). Under the assumption that the dielectric medium is lossy
without net gain, the discontinuity of $\bchi({\bf r},{\bf r}',\omega)$ for
$\omega>0$ is a positive-definite integral kernel. Hence, one may use
(\ref{4.3}) to introduce a tensor $\bfsfT({\bf r},{\bf r}',\omega)$, which
is uniquely defined up to a unitary transformation. Subsequently, one can
construct the model Hamiltonian (\ref{2.21}) and proceed with its
diagonalization. In conclusion, our formalism applies to all anisotropic
lossy dielectric media with a spatio-temporal dispersion that is compatible
with the fundamental principles of causality and positive-definiteness of
the dissipative energy loss. Incidentally, it may be remarked that
amplifying dielectric media, which have been treated in the context of the
phenomenological quantization scheme as well \cite{KSW01,MLABJ97}, are not
covered by the present damped-polariton model. To describe media with a
sustained gain, e.g.\ a laser above threshold, one has to incorporate a
driving mechanism in the Hamiltonian, which accounts for the ongoing
input of energy that is indispensable for a stationary gain.

As we have shown, the time evolution of the dynamical variables for field
and matter can be determined completely by deriving the operators that
diagonalize the Hamiltonian. The diagonalizing operators are closely
related to the noise part of the polarization density, which plays an
important role in the phenomenological quantization scheme. The proof of
the commutation properties of the noise polarization density follows from
its relation to the diagonalizing operators.

In setting up our model Hamiltonian we have avoided to introduce a bath
environment from the beginning. The subsequent formalism could be developed
without ever discussing such a bath. Nevertheless, one may be interested in
an analysis of the complete set of degrees of freedom of the dielectric
medium in our model. If that analysis is carried out, one finds, as we have
seen above, that specific combinations of medium variables can be
associated to what may be called a bath. The coupling of the polarization
to this bath can be held responsible for the dissipative losses
that characterize a dispersive dielectric.

\ack
I would like to thank dr.~A.J.~van Wonderen for numerous discussions and
critical comments. 

\appendix
\section{Lagrangian formulation}
\setcounter{section}{1}
In this appendix we shall show how the Hamiltonian (\ref{2.21}) can be related
to a Lagrange formalism. We start by postulating the following Lagrangian for
an anisotropic linear dielectric with spatio-temporal dispersion that
interacts with the electromagnetic field:
\begin{eqnarray}
 L=\int d{\bf r} \left\{\half\, \varepsilon_0\,[\dot{\bf A}({\bf r})]^2
-\frac{1}{2\mu_0}[\bnabla\wedge{\bf A}({\bf r})]^2\right\}\nonumber\\
+\half\int d{\bf r} \int_0^\infty d\omega\, 
\left\{[\dot{\bf Q}_m({\bf r},\omega)]^2
-\omega^2\, [{\bf Q}_m({\bf r},\omega)]^2\right\}\nonumber\\
+\int d{\bf r}\, \dot{\bf P}({\bf r})\cdot{\bf A}({\bf r})
- \frac{1}{2\varepsilon_0}\int d{\bf r} \, \left\{[{\bf P}({\bf r})]_L\right\}^2.
\label{A.1}
\end{eqnarray}
Here ${\bf A}({\bf r})$ is the transverse vector potential and ${\bf
Q}_m({\bf r},\omega)$ are material coordinates depending on position and
frequency. The polarization density ${\bf P}({\bf r})$ is taken to be an
anisotropic and non-local linear combination of these material coordinates
of the form
\begin{equation}
{\bf P}({\bf r})=\int d{\bf r}'\int_0^\infty d\omega' \, {\bf Q}_m({\bf
  r}',\omega')\cdot \bfsfT_0({\bf r}',{\bf r},\omega')
\label{A.2}
\end{equation} 
with a real tensor coefficient $\bfsfT_0({\bf r},{\bf r}',\omega)$. One
easily verifies that the Lagrangian equations have the form
\begin{eqnarray}
\Delta {\bf A}({\bf r},t)-\frac{1}{c^2}\, \ddot{\bf A}({\bf r},t)=
-\mu_0\, \left[\dot{\bf P}({\bf r},t)\right]_T
\label{A.3}\\
\ddot{\bf Q}_m({\bf r},\omega,t)+\omega^2\, {\bf Q}_m({\bf r},\omega,t)=
\int d{\bf r}\, \bfsfT_0({\bf r},{\bf r}',\omega)\cdot{\bf E}({\bf r}',t)
\label{A.4}
\end{eqnarray}
with the electric field given as ${\bf E}({\bf r},t) =-\dot{\bf A}({\bf
r},t)-(1/\varepsilon_0)\, [{\bf P}({\bf r},t)]_L$. The first Lagrangian
differential equation is consistent with Maxwell's equation, as it
should. The second Lagrangian equation shows that the material coordinates
are harmonic variables that are driven by the electric field in an
anisotropic and non-local way.

Introducing the momenta $\bPi({\bf
r})$ and ${\bf P}_m({\bf r},\omega)$ associated to ${\bf A}$ and ${\bf
Q}_m$ as
\begin{eqnarray}
\bPi({\bf r})=\varepsilon_0\, \dot{\bf A}({\bf r})
\label{A.5}\\
{\bf P}_m({\bf r},\omega)= \dot{\bf Q}_m({\bf r},\omega)+
\int d{\bf r}'\, \bfsfT_0({\bf r},{\bf r}',\omega)\cdot{\bf A}({\bf r}')
\label{A.6}
\end{eqnarray}
we obtain the Hamiltonian corresponding to (\ref{A.1}) in the standard
fashion. The result is:
\begin{eqnarray}
\fl H=\int d{\bf r} \left\{\frac{1}{2\varepsilon_0}
[\bPi({\bf r})]^2+\frac{1}{2\mu_0}[\bnabla\wedge{\bf A}({\bf r})]^2\right\}
\nonumber\\
+\half \int d{\bf r}\int_0^\infty d\omega \, 
\left\{ [{\bf P}_m({\bf r},\omega)]^2+
\omega^2\, [{\bf Q}_m({\bf r},\omega)]^2\right\}\nonumber\\
-\int d{\bf r}\int d{\bf r}'\int_0^\infty d\omega\, 
{\bf P}_m({\bf r},\omega)\cdot\bfsfT_0({\bf r},{\bf r}',\omega)\cdot{\bf
  A}({\bf r}')\nonumber\\
+\half \int d{\bf r}\int d{\bf r}'\int d{\bf r}'' \int_0^\infty d\omega\, 
{\bf A}({\bf r})\cdot \tilde{\bfsfT_0}({\bf r}',{\bf r},\omega)\cdot
\bfsfT_0({\bf r}',{\bf r}'',\omega)\cdot{\bf A}({\bf r}'')\nonumber\\
+ \frac{1}{2\varepsilon_0}\int d{\bf r} \, \left\{[{\bf P}({\bf r})]_L\right\}^2.
\label{A.7}
\end{eqnarray}

As a final step we wish to rewrite the Hamiltonian in terms of material
creation and annihilation operators ${\bf C}^\dagger_m$ and ${\bf C}_m$ as
used in (\ref{2.21}) of the main text. We introduce the latter by writing
\begin{eqnarray}
\fl{\bf P}_m({\bf r},\omega)=\left(\frac{\hbarit\omega}{2}\right)^{1/2}\,
\int d{\bf r}'\, \left[
{\bf C}_m({\bf r}',\omega)\cdot\bfsfU({\bf r}',{\bf r},\omega)
+ {\bf C}^\dagger_m({\bf r}',\omega)\cdot\bfsfU^\ast({\bf r}',{\bf
  r},\omega)\right]
\label{A.8}\\
\fl{\bf Q}_m({\bf r},\omega)=\rmi\,\left(\frac{\hbarit}{2\omega}\right)^{1/2}\,
\int d{\bf r}'\, \left[
{\bf C}_m({\bf r}',\omega)\cdot\bfsfU({\bf r}',{\bf r},\omega)
- {\bf C}^\dagger_m({\bf r}',\omega)\cdot\bfsfU^\ast({\bf r}',{\bf
  r},\omega)\right]
\label{A.9}
\end{eqnarray}
with tensorial coefficients $\bfsfU$ that satisfy the unitarity condition
\begin{equation}
\int d{\bf r}''\, \tilde{\bfsfU}({\bf r}'',{\bf r},\omega)\cdot
\bfsfU^\ast({\bf r}'',{\bf r}',\omega)=\bfsfI\, \delta({\bf r}-{\bf r}').
\label{A.10}
\end{equation}
Substituting (\ref{A.8}) and (\ref{A.9}) in (\ref{A.7}) we recover
(\ref{2.21}) of the main text, with $\bfsfT$ given by
\begin{equation}
\bfsfT({\bf r},{\bf r}',\omega)=-(2\hbarit\omega)^{-1/2}\,
\int d{\bf r}''\, \bfsfU({\bf r},{\bf r}'',\omega)\cdot
\bfsfT_0({\bf r}'',{\bf r}',\omega).
\label{A.11}
\end{equation}
Since $\bfsfT_0$ is real, one finds that introducing $\bfsfT$ in this way
implies that it fulfills the relation
\begin{equation}
\int d{\bf r}''\, \tilde{\bfsfT}({\bf r}'',{\bf r},\omega)\cdot
\bfsfT^\ast({\bf r}'',{\bf r}',\omega)- {\rm c.c.}=0
\label{A.12}
\end{equation}
which is consistent with (but somewhat stronger than) conditions
(\ref{2.8}) and (\ref{2.22}) imposed on $\bfsfT$ in the main text. Adopting
the above stronger relation implies that the susceptibility (\ref{3.12})
acquires an additional symmetry property on a par with (\ref{4.1}) and
(\ref{4.2}), namely $\bchi({\bf r}, {\bf r}',z)=\bchi({\bf r},{\bf
r}',-z)$. As the validity of (\ref{A.12}) is not essential in setting up
our Hamilton formalism, we have refrained from using it in the main text.

\section{Evaluation of the tensorial coefficients $\bfsff_i$}

In this appendix we will show how the equations (\ref{3.3}-\ref{3.6}) can
be solved. We start by using (\ref{3.3}) to eliminate $\bfsff_1$ from
(\ref{3.4}). As a result we obtain the differential equation:
\begin{eqnarray}
\fl \Delta' \bfsff_2({\bf r},{\bf r}',\omega)
+\frac{\omega^2}{c^2}\, \bfsff_2({\bf r},{\bf r}',\omega)-\mu_0\,\hbarit
\int d{\bf r}''\, \bfsff_2({\bf r},{\bf r}'',\omega)
\cdot\left[\bfsfF({\bf r}'',{\bf r}')\right]_{T'}\nonumber\\
\fl -\rmi\,\mu_0
\int d{\bf r}''\int_0^\infty d\omega''\, \omega''\left\{
\bfsff_3({\bf r},{\bf r}'',\omega,\omega'')\cdot
[\bfsfT^\ast({\bf r}'',{\bf r}',\omega'')]_{T'}\right.\nonumber\\
 \left. -\bfsff_4({\bf r},{\bf r}'',\omega,\omega'')\cdot
[\bfsfT({\bf r}'',{\bf r}',\omega'')]_{T'}\right\}=0.
\label{B.1}
\end{eqnarray}
To get an expression for the last two terms we use (\ref{3.5}) and
(\ref{3.6}). First, we multiply (\ref{3.5}) by $\bfsfT^\ast({\bf r}',{\bf
r}'''',\omega')$ and integrate over ${\bf r}'$. Relabeling the dummy
variables we get
\begin{eqnarray}
\fl -\rmi\,\hbarit\, \omega'\int d{\bf r}''\int d{\bf r}'''\, 
\bfsff_2({\bf r},{\bf r}'',\omega)
\cdot \tilde{\bfsfT}({\bf r}''',{\bf r},'',\omega')\cdot
\bfsfT^\ast({\bf r}''',{\bf r}',\omega')\nonumber\\
\fl -(\omega-\omega')\int d{\bf r}''\, 
\bfsff_3({\bf r},{\bf r}'',\omega,\omega')\cdot
\bfsfT^\ast({\bf r}'',{\bf r}',\omega')\nonumber\\
\fl +\frac{\hbarit}{\varepsilon_0}\int d{\bf r}''\int d{\bf r}''' \int 
d{\bf r}''''\int_0^\infty d\omega'' \left\{
\bfsff_3({\bf r},{\bf r}'',\omega,\omega'')\cdot
[\bfsfT^\ast({\bf r}'',{\bf r}''',\omega'')]_{L'''}\right.\nonumber\\
\fl\left. +\bfsff_4({\bf r},{\bf r}'',\omega,\omega'')\cdot
[\bfsfT({\bf r}'',{\bf r}''',\omega'')]_{L'''}\right\}\cdot
\tilde{\bfsfT}({\bf r}'''',{\bf r}''',\omega')\cdot
\bfsfT^\ast({\bf r}'''',{\bf r}',\omega')=0.
\label{B.2}
\end{eqnarray}
A similar relation is obtained by multiplying (\ref{3.6}) by $\bfsfT({\bf
  r}',{\bf r}'''',\omega')$ and integrating over ${\bf r}'$:  
\begin{eqnarray}
\fl -\rmi\,\hbarit\, \omega'\int d{\bf r}''\int d{\bf r}'''\, 
\bfsff_2({\bf r},{\bf r}'',\omega)
\cdot \tilde{\bfsfT}^\ast({\bf r}''',{\bf r},'',\omega')\cdot
\bfsfT({\bf r}''',{\bf r}',\omega')\nonumber\\
\fl -(\omega+\omega')\int d{\bf r}''\, 
\bfsff_4({\bf r},{\bf r}'',\omega,\omega')\cdot
\bfsfT({\bf r}'',{\bf r}',\omega')\nonumber\\
\fl -\frac{\hbarit}{\varepsilon_0}\int d{\bf r}''\int d{\bf r}''' \int 
d{\bf r}''''\int_0^\infty d\omega'' \left\{
\bfsff_3({\bf r},{\bf r}'',\omega,\omega'')\cdot
[\bfsfT^\ast({\bf r}'',{\bf r}''',\omega'')]_{L'''}\right.\nonumber\\
\fl\left. +\bfsff_4({\bf r},{\bf r}'',\omega,\omega'')\cdot
[\bfsfT({\bf r}'',{\bf r}''',\omega'')]_{L'''}\right\}\cdot
\tilde{\bfsfT}^\ast({\bf r}'''',{\bf r}''',\omega')\cdot
\bfsfT({\bf r}'''',{\bf r}',\omega')=0.
\label{B.3}
\end{eqnarray}
We add (\ref{B.2}) and (\ref{B.3}). Upon integrating over $\omega'$ and using
(\ref{2.8}) and (\ref{2.23}) we get
\begin{eqnarray}
\fl -\rmi\,\hbarit\int d{\bf r}''\, \bfsff_2({\bf r},{\bf r}'',\omega)\cdot
\bfsfF({\bf r}'',{\bf r}')\nonumber\\
\fl -\omega \int d{\bf r}''\int_0^\infty d\omega''\, 
\left[\bfsff_3({\bf r},{\bf r}'',\omega,\omega'')\cdot
\bfsfT^\ast({\bf r}'',{\bf r}',\omega'')
+\bfsff_4({\bf r},{\bf r}'',\omega,\omega'')\cdot
\bfsfT({\bf r}'',{\bf r}',\omega'')\right]\nonumber\\
\fl +\int d{\bf r}''\int_0^\infty d\omega''\,\omega''\, 
\left[\bfsff_3({\bf r},{\bf r}'',\omega,\omega'')\cdot
\bfsfT^\ast({\bf r}'',{\bf r}',\omega'')
-\bfsff_4({\bf r},{\bf r}'',\omega,\omega'')\cdot
\bfsfT({\bf r}'',{\bf r}',\omega'')\right]=0.\nonumber\\
\label{B.4}
\end{eqnarray}
By taking the transverse part of this relation with respect to ${\bf r}'$
we get an identity that can be used to rewrite (\ref{B.1}) in the form:
\begin{eqnarray}
 -\left[\bfsff_2({\bf r},{\bf r}',\omega)\times \overleftarrow{\bnabla}'\right]
\times \overleftarrow{\bnabla}'
+\frac{\omega^2}{c^2}\, \bfsff_2({\bf r},{\bf r}',\omega)\nonumber\\
 -\rmi\,\mu_0\, \omega
\int d{\bf r}''\int_0^\infty d\omega''\left\{
\bfsff_3({\bf r},{\bf r}'',\omega,\omega'')\cdot
[\bfsfT^\ast({\bf r}'',{\bf r}',\omega'')]_{T'}\right.\nonumber\\
 \left. +\bfsff_4({\bf r},{\bf r}'',\omega,\omega'')\cdot
[\bfsfT({\bf r}'',{\bf r}',\omega'')]_{T'}\right\}=0.
\label{B.5}
\end{eqnarray}
Here we used the transversality of $\bfsff_2({\bf r},{\bf r}',\omega)$ in
its second argument to write the first term as a repeated vector product, with
the spatial derivative operator $\bnabla'$ acting to the left on the argument
${\bf r}'$ of the function $\bfsff_2({\bf r},{\bf r}',\omega)$.  

The integral in (\ref{B.5}) contains the transverse parts of $\bfsfT$ and
$\bfsfT^\ast$ only. A more natural form of the differential equation, with
the full tensors $\bfsfT$ and $\bfsfT^\ast$, is obtained by introducing
instead of $\bfsff_2$ a new tensor $\bfsfg$ defined as:
\begin{eqnarray}
\bfsfg({\bf r},{\bf r}',\omega)\equiv 
\rmi\,\omega\,\bfsff_2({\bf r},{\bf r}',\omega) \nonumber\\
 -\frac{1}{\varepsilon_0}
\int d{\bf r}''\int_0^\infty d\omega''\left\{
\bfsff_3({\bf r},{\bf r}'',\omega,\omega'')\cdot
[\bfsfT^\ast({\bf r}'',{\bf r}',\omega'')]_{L'}\right.\nonumber\\
 \left. +\bfsff_4({\bf r},{\bf r}'',\omega,\omega'')\cdot
[\bfsfT({\bf r}'',{\bf r}',\omega'')]_{L'}\right\}.
\label{B.6}
\end{eqnarray}
It satisfies a differential equation, which follows from (\ref{B.5}) as:
\begin{eqnarray}
-\left[\bfsfg({\bf r},{\bf r}',\omega)\times \overleftarrow{\bnabla}'\right]
\times \overleftarrow{\bnabla}'
+\frac{\omega^2}{c^2}\, \bfsfg({\bf r},{\bf r}',\omega)\nonumber\\
+\mu_0\, \omega^2
\int d{\bf r}''\int_0^\infty d\omega''\left[
\bfsff_3({\bf r},{\bf r}'',\omega,\omega'')\cdot
\bfsfT^\ast({\bf r}'',{\bf r}',\omega'')
\right.\nonumber\\
 \left. 
+\bfsff_4({\bf r},{\bf r}'',\omega,\omega'')\cdot
\bfsfT({\bf r}'',{\bf r}',\omega'')\right]=0.
\label{B.7}
\end{eqnarray}
The integral contribution still depends on $\bfsff_3$ and $\bfsff_4$, so
that the differential equation is not yet in closed form. However, we may
rewrite the integral in such a way that its relation to $\bfsfg$ becomes
obvious. This can be achieved with the help of the identity:
\begin{eqnarray}
\fl \int d{\bf r}''\int_0^\infty d\omega''
\left[\bfsff_3({\bf r},{\bf r}'',\omega,\omega'')\cdot
\bfsfT^\ast({\bf r}'',{\bf r}',\omega'')
+\bfsff_4({\bf r},{\bf r}'',\omega,\omega'')\cdot
\bfsfT({\bf r}'',{\bf r}',\omega'')\right]=\nonumber\\
 =\varepsilon_0\int d{\bf r}''\, \bfsfg({\bf r},{\bf r}'',\omega)\cdot
\bchi({\bf r}'',{\bf r}',\omega-\rmi\,0)+\bfsfs({\bf r},{\bf r}',\omega).
\label{B.8}
\end{eqnarray}
which contains a tensor $\bfsfs({\bf r},{\bf r}',\omega)$ that arises while
avoiding a pole in the complex frequency plane, as we shall see
below. Furthermore the right-hand side contains the susceptibility tensor
$\bchi$ that has been defined in (\ref{3.12}). In (\ref{B.8}) the
frequency is chosen to be in the lower half of the complex plane just below
the real axis. Correspondingly, the term $-\rm i \, 0$ is an
infinitesimally small number on the negative imaginary axis.

To prove (\ref{B.8}) we divide (\ref{B.2}) by $\omega'-\omega+\rmi\, 0$,
with $\rmi\, 0$ an infinitesimally small imaginary number. The result is:
\begin{eqnarray}
\fl -\rmi \, \hbarit\, \frac{\omega'}{\omega'-\omega+\rmi\, 0}\int d{\bf
 r}''\int d{\bf r}'''\, \bfsff_2({\bf r},{\bf r}'',\omega)\cdot
\tilde{\bfsfT}({\bf r}''',{\bf r}'',\omega')\cdot
\bfsfT^\ast({\bf r}''',{\bf r}',\omega')\nonumber\\
\fl +\int d{\bf r}''\, \bfsff_3({\bf r},{\bf r}'',\omega,\omega')\cdot
\bfsfT^\ast({\bf r}'',{\bf r}',\omega')\nonumber\\
\fl +\frac{\hbarit}{\varepsilon_0}\, 
\frac{1}{\omega'-\omega+\rmi\, 0}\int d{\bf r}''\int d{\bf r}''' 
\int d{\bf r}''''\int_0^\infty d\omega'' \left\{
\bfsff_3({\bf r},{\bf r}'',\omega,\omega'')\cdot
\left[\bfsfT^\ast({\bf r}'',{\bf r}''',\omega'')\right]_{L'''}\right.\nonumber\\
\fl\left. +\bfsff_4({\bf r},{\bf r}'',\omega,\omega'')\cdot
\left[\bfsfT({\bf r}'',{\bf r}''',\omega'')\right]_{L'''}\right\}
\cdot\tilde{\bfsfT}({\bf r}'''',{\bf r}''',\omega')\cdot
\bfsfT^\ast({\bf r}'''',{\bf r}',\omega')\nonumber\\
=\delta(\omega-\omega')\, \bfsfs({\bf r},{\bf r}',\omega).
\label{B.9}
\end{eqnarray}
At the right-hand side, we have introduced a term proportional to the
delta function $\delta(\omega-\omega')$ to account for the fact that the
division by $\omega'-\omega+\rmi\, 0$ yields a singular result for
$\omega=\omega'$, as discussed in \cite{F61}. The coefficient $\bfsfs$ is as yet unknown.
Likewise, upon dividing (\ref{B.3}) by $\omega+\omega'$ we obtain:
\begin{eqnarray}
\fl -\rmi \, \hbarit\, \frac{\omega'}{\omega+\omega'}\int d{\bf r}''\int
 d{\bf r}'''\, \bfsff_2({\bf r},{\bf r}'',\omega)\cdot
 \tilde{\bfsfT}^\ast({\bf r}''',{\bf r}'',\omega')\cdot \bfsfT({\bf
 r}''',{\bf r}',\omega')\nonumber\\ \fl -\int d{\bf r}''\, \bfsff_4({\bf
 r},{\bf r}'',\omega,\omega')\cdot \bfsfT({\bf r}'',{\bf
 r}',\omega')\nonumber\\ \fl -\frac{\hbarit}{\varepsilon_0}\,
 \frac{1}{\omega+\omega'}\int d{\bf r}''\int d{\bf r}''' \int d{\bf
 r}''''\int_0^\infty d\omega'' \left\{ \bfsff_3({\bf r},{\bf
 r}'',\omega,\omega'')\cdot \left[\bfsfT^\ast({\bf r}'',{\bf
 r}''',\omega'')\right]_{L'''}\right.\nonumber\\ \fl\left. +\bfsff_4({\bf
 r},{\bf r}'',\omega,\omega'')\cdot \left[\bfsfT({\bf r}'',{\bf
 r}''',\omega'')\right]_{L'''}\right\} \cdot\tilde{\bfsfT}^\ast({\bf
 r}'''',{\bf r}''',\omega')\cdot \bfsfT({\bf r}'''',{\bf r}',\omega')=0.
\label{B.10}
\end{eqnarray}
 We subtract (\ref{B.10}) from (\ref{B.9}) and integrate over
$\omega'$. Inspecting the contributions from the last terms at the
left-hand sides we find it useful to introduce the susceptibility tensor
(\ref{3.12}).  Upon employing moreover (\ref{2.8}), we get as a result of
combining (\ref{B.9}) and (\ref{B.10}):
\begin{eqnarray}
\fl -\rmi\, \varepsilon_0\, \omega\int d{\bf r}''\, 
\bfsff_2({\bf r},{\bf r}'',\omega)\cdot 
\bchi({\bf r}'',{\bf r}',\omega-\rmi\,0)\nonumber\\
\fl + \int d{\bf r}''\int_0^\infty d\omega''
\left[\bfsff_3({\bf r},{\bf r}'',\omega,\omega'')\cdot
\bfsfT^\ast({\bf r}'',{\bf r}',\omega'')
+\bfsff_4({\bf r},{\bf r}'',\omega,\omega'')\cdot
\bfsfT({\bf r}'',{\bf r}',\omega'')\right]\nonumber\\
\fl +\int d{\bf r}''\int d{\bf r}'''\int_0^\infty d\omega''
\left\{
\bfsff_3({\bf r},{\bf r}'',\omega,\omega'')\cdot
\left[\bfsfT^\ast({\bf r}'',{\bf r}''',\omega'')\right]_{L'''}\right.\nonumber\\
\fl \left. +\bfsff_4({\bf r},{\bf r}'',\omega,\omega'')\cdot
\left[\bfsfT({\bf r}'',{\bf r}''',\omega'')\right]_{L'''}\right\}
\cdot \bchi({\bf r}''',{\bf
  r}',\omega-\rmi\,0)
=\bfsfs({\bf r},{\bf r}',\omega).
\label{B.11}
\end{eqnarray}
When (\ref{B.6}) is invoked to eliminate $\bfsff_2$ in favour of $\bfsfg$,
we recover (\ref{B.8}).

Having established (\ref{B.8}) we insert it in (\ref{B.7}) so as to
arrive at an inhomogeneous wave equation for $\bfsfg$:
\begin{eqnarray}
 -\left[\bfsfg({\bf r},{\bf r}',\omega)\times \overleftarrow{\bnabla}'\right]
\times \overleftarrow{\bnabla}'
+\frac{\omega^2}{c^2}\, \bfsfg({\bf r},{\bf r}',\omega)\nonumber\\
 +\frac{\omega^2}{c^2}\int d{\bf r}''\, \bfsfg({\bf r},{\bf r}'',\omega)\cdot
\bchi({\bf r}'',{\bf r}',\omega-\rmi\,0)=-\mu_0\,
\omega^2\, \bfsfs({\bf r},{\bf r}',\omega).
\label{B.12}
\end{eqnarray}
To solve $\bfsfg$ from this differential equation we employ the Green
function $\bfsfG({\bf r},{\bf r}',z)$ associated to the operator at the
left-hand side. It has been defined in (\ref{3.11}). In terms of this Green
function the solution of (\ref{B.12}) reads:
\begin{equation}
\bfsfg({\bf r},{\bf r}',\omega)=
-\mu_0\, \omega^2\int d{\bf r}''\, \bfsfs({\bf r},{\bf r}'',\omega)\cdot
\bfsfG({\bf r}'',{\bf r}',\omega-\rmi\, 0).
\label{B.13}
\end{equation}

With the use of the above expression for $\bfsfg$ in terms of $\bfsfs$, we
can evaluate the coefficients $\bfsff_i$ successively. Let us start with
$\bfsff_2$. From (\ref{B.7}) it follows that the integral contribution in
(\ref{B.6}) is proportional to the longitudinal part $[\bfsfg({\bf r},{\bf
r}',\omega)]_{L'}$ of $\bfsfg$. As a consequence, (\ref{B.6}) implies that
$\bfsff_2$ equals $-(\rmi/\omega)\, [\bfsfg({\bf r},{\bf
r}',\omega)]_{T'}$. Hence, upon using (\ref{B.13}) and introducing the
tensor $\bfsfu$ by writing $\bfsfs$ as
\begin{equation}
\bfsfs({\bf r},{\bf r}',\omega)=\int d{\bf r}''\, 
\bfsfu({\bf r},{\bf r}'',\omega)\cdot
\bfsfT^\ast({\bf r}'',{\bf r}',\omega)
\label{B.14}
\end{equation}
we obtain: 
\begin{equation}
\fl \bfsff_2({\bf r},{\bf r}',\omega)=\rmi\, \mu_0\, \omega
\int d{\bf r}'' \int d{\bf r}'''\, \bfsfu({\bf r},{\bf r}'',\omega)\cdot
\bfsfT^\ast({\bf r}'',{\bf r}''',\omega)\cdot
[\bfsfG({\bf r}''',{\bf r}',\omega-\rmi\, 0)]_{T'}.
\label{B.15}
\end{equation}
On a par with this expression we get from (\ref{3.3}):
\begin{equation}
\fl \bfsff_1({\bf r},{\bf r}',\omega)=\frac{\omega^2}{c^2}
\int d{\bf r}'' \int d{\bf r}'''\, \bfsfu({\bf r},{\bf r}'',\omega)\cdot
\bfsfT^\ast({\bf r}'',{\bf r}''',\omega)\cdot
[\bfsfG({\bf r}''',{\bf r}',\omega-\rmi\, 0)]_{T'}.
\label{B.16}
\end{equation}

Expressions for $\bfsff_3$ and $\bfsff_4$ follow from (\ref{B.9}) and
(\ref{B.10}) upon using (\ref{B.13})--(\ref{B.15}) and the longitudinal part
of (\ref{B.7}). We get:
\begin{eqnarray}
\fl \bfsff_3({\bf r},{\bf r}',\omega,\omega')=\delta(\omega-\omega')\, \,
\bfsfu({\bf r},{\bf r}',\omega) \nonumber\\
 -\mu_0\,\hbarit\,\omega \int d{\bf r}''\int d{\bf r}''' 
\int d{\bf r}''''\, \bfsfu({\bf r},{\bf r}'',\omega)\cdot
\bfsfT^\ast({\bf r}'',{\bf r}''',\omega) \nonumber\\
\cdot [\bfsfG({\bf r}''',{\bf r}'''',\omega-\rmi \, 0)]_{T''''}\cdot
\tilde{\bfsfT}({\bf r}',{\bf r}'''',\omega')\nonumber\\
 +\mu_0\,\hbarit\,\frac{\omega^2}{\omega-\omega'-\rmi\, 0}
\int d{\bf r}''\int d{\bf r}''' \int d{\bf r}''''\, 
\bfsfu({\bf r},{\bf r}'',\omega)\cdot
\bfsfT^\ast({\bf r}'',{\bf r}''',\omega)\nonumber\\
\cdot \bfsfG({\bf r}''',{\bf r}'''',\omega-\rmi\, 0) \cdot
\tilde{\bfsfT}({\bf r}',{\bf r}'''',\omega')
\label{B.17}\\
\fl \bfsff_4({\bf r},{\bf r}',\omega,\omega')=
\mu_0\,\hbarit\,\omega \int d{\bf r}''\int d{\bf r}''' 
\int d{\bf r}''''\, \bfsfu({\bf r},{\bf r}'',\omega)\cdot
\bfsfT^\ast({\bf r}'',{\bf r}''',\omega) \nonumber\\
\cdot [\bfsfG({\bf r}''',{\bf r}'''',\omega-\rmi \, 0)]_{T''''}\cdot
\tilde{\bfsfT}^\ast({\bf r}',{\bf r}'''',\omega')\nonumber\\
 -\mu_0\,\hbarit\,\frac{\omega^2}{\omega+\omega'}
\int d{\bf r}''\int d{\bf r}''' \int d{\bf r}''''\, 
\bfsfu({\bf r},{\bf r}'',\omega)\cdot
\bfsfT^\ast({\bf r}'',{\bf r}''',\omega)\nonumber\\
\cdot \bfsfG({\bf r}''',{\bf r}'''',\omega-\rmi\, 0) \cdot
\tilde{\bfsfT}^\ast({\bf r}',{\bf r}'''',\omega').
\label{B.18}
\end{eqnarray}
The introduction of $\bfsfu$ instead of $\bfsfs$ has led to the simple form
of the first term at the right-hand side of (\ref{B.17}). 

\section{The tensor $\bfsfu$}

In appendix B the coefficients $\bfsff_i$ have been obtained. They are all
proportional to the tensor $\bfsfu({\bf r},{\bf r}',\omega)$. In this
section, we shall show how this tensor can be determined. 

After insertion of the coefficients in the expression (\ref{3.2}) it
follows that the diagonalizing operator ${\bf C}({\bf r},\omega)$ itself is
proportional to $\bfsfu$ as well. Since ${\bf C}({\bf r},\omega)$ and its
Hermitian conjugate must satisfy canonical commutation relations of the
general form (\ref{2.4}), $\bfsfu$ has to fulfil a constraint that is
obtained by evaluating the commutator $[{\bf C}({\bf r},\omega), {\bf
C}^\dagger({\bf r}',\omega')]$. In fact, substituting the expression
(\ref{3.2}) and its Hermitian conjugate in the commutator and employing
(\ref{2.2}) and (\ref{2.4}) we arrive at the condition
\begin{eqnarray}
\fl \rmi\, \hbarit\int d{\bf r}''\, \left[
\bfsff_1({\bf r},{\bf r}'',\omega) 
\cdot \tilde{\bfsff}^\ast_2({\bf r}',{\bf r}'',\omega')-
\bfsff_2({\bf r},{\bf r}'',\omega) 
\cdot \tilde{\bfsff}^\ast_1({\bf r}',{\bf r}'',\omega')\right] \nonumber\\
\fl +\int d{\bf r}''\int_0^\infty d\omega''\, 
\left[
\bfsff_3({\bf r},{\bf r}'',\omega,\omega'') 
\cdot \tilde{\bfsff}^\ast_3({\bf r}',{\bf r}'',\omega',\omega'')\right.\nonumber\\
\left.-\bfsff_4({\bf r},{\bf r}'',\omega,\omega'') 
\cdot \tilde{\bfsff}^\ast_4({\bf r}',{\bf
  r}'',\omega',\omega'')\right]=
\bfsfI \,  \delta({\bf r}-{\bf r}')\, \delta(\omega-\omega').
\label{C.1}
\end{eqnarray}
After the insertion of the formulae for $\bfsff_i$ we arrive at a bilinear
condition for $\bfsfu$ of the form:
\begin{equation}
\fl \int d{\bf r}'' \int d{\bf r}''' \, \bfsfu({\bf r},{\bf r}'',\omega)\cdot
\bfsfM({\bf r}'',{\bf r}''',\omega,\omega')\cdot
\tilde{\bfsfu}^\ast({\bf r}',{\bf r}''',\omega')=
\bfsfI \,  \delta({\bf r}-{\bf r}')\, \delta(\omega-\omega').
\label{C.2}
\end{equation}

The explicit form of the tensorial integral kernel $\bfsfM$ follows by
evaluating (\ref{C.1}) with (\ref{B.15})--(\ref{B.18}). It contains
contributions with a variable number of Green functions $\bfsfG$. The
simplest contribution $\bfsfM_0$ is that without a Green function, which
is found to be
\begin{equation}
\bfsfM_0({\bf r},{\bf r}',\omega,\omega')=
\bfsfI \,  \delta({\bf r}-{\bf r}')\, \delta(\omega-\omega').
\label{C.3}
\end{equation}

The next contribution $\bfsfM_1$ consists of all terms containing a single
Green function:
\begin{eqnarray}
\fl \bfsfM_1({\bf r},{\bf r}',\omega,\omega')=
-\mu_0\,\hbarit\int d{\bf r}''\int d{\bf r}'''\, 
\bfsfT^\ast({\bf r},{\bf r}'',\omega)\cdot\left\{\rule{0mm}{5mm}
\omega \left[ \bfsfG({\bf r}'',{\bf r}''',\omega-\rmi\,0)\right]_{T'''}\right.
\nonumber\\
 +\omega' \left[
\tilde{\bfsfG}^\ast({\bf r}''',{\bf r}'',\omega'-\rmi\,0)\right]_{T''}
-\frac{\omega^2}{\omega-\omega'-\rmi\, 0}\, 
\bfsfG({\bf r}'',{\bf r}''',\omega-\rmi\,0)\nonumber\\
 \left.+\frac{\omega'^2}{\omega-\omega'-\rmi\, 0}\, 
\tilde{\bfsfG}^\ast({\bf r}''',{\bf r}'',\omega'-\rmi\,0)\right\}
\cdot \tilde{\bfsfT}({\bf r}',{\bf r}''',\omega').
\label{C.4}
\end{eqnarray}
The first two terms contain the transverse part of the Green function,
whereas the last two depend on the full Green function. 

Finally, we have got the contributions with two Green functions. Since
again both the full Green function and its transverse part show up, we can
distinguish various types of terms. The terms with two transverse Green
functions become upon invoking (\ref{2.8}):
\begin{eqnarray}
\fl \bfsfM_{2TT}({\bf r},{\bf r}',\omega,\omega')=
\frac{\mu_0\, \hbarit}{c^2}\, \omega\, \omega'(\omega+\omega')
\int d{\bf r}''\int d{\bf r}''' \int d{\bf r}''''\, 
\bfsfT^\ast({\bf r},{\bf r}'',\omega)\nonumber\\
\fl \cdot \left[\bfsfG({\bf r}'',{\bf r}''',\omega -\rmi\, 0)\right]_{T'''}
 \cdot  \left[\tilde{\bfsfG}^\ast({\bf r}'''',{\bf r}''',\omega' -\rmi\, 0)
\right]_{T'''}\cdot \tilde{\bfsfT}({\bf r}',{\bf r}'''',\omega').
\label{C.5}
\end{eqnarray}
The terms with one transverse and one full Green function are
\begin{eqnarray}
\fl \bfsfM_{2T}({\bf r},{\bf r}',\omega,\omega')=
\frac{\mu_0\, \hbarit}{c^2}\, \omega\, \omega' 
\int d{\bf r}''\int d{\bf r}'''\int d{\bf r}''''\int d{\bf r}^v \,
\bfsfT^\ast({\bf r},{\bf r}'',\omega)\nonumber\\
\fl \cdot\left\{ 
\omega\,\bfsfG({\bf r}'',{\bf r}''',\omega-\rmi\,  0)
\cdot \bchi({\bf r}''',{\bf r}'''',\omega-\rmi\, 0)
\cdot \left[
\tilde{\bfsfG}^\ast({\bf r}^v,{\bf r}'''',\omega'-\rmi\, 0)\right]_{T''''}
\right.\nonumber\\
\fl \left. +\omega'\,\left[ \bfsfG({\bf r}'',{\bf r}''',\omega-\rmi\,  0)\right]_{T'''}
\cdot \tilde{\bchi}^\ast({\bf r}'''',{\bf r}''',\omega'-\rmi\, 0)
\cdot \tilde{\bfsfG}^\ast({\bf r}^v,{\bf r}'''',\omega'-\rmi\, 0)\right\}
\cdot \tilde{\bfsfT}({\bf r}',{\bf r}^v,\omega')\nonumber\\
\label{C.6}
\end{eqnarray}
where we introduced the susceptibility tensor (\ref{3.12}). The last
set of terms we have to consider are those with two full Green
functions. Again using (\ref{3.12}) we get 
\begin{eqnarray}
\fl \bfsfM_{2}({\bf r},{\bf r}',\omega,\omega')=
\frac{\mu_0\,\hbarit}{c^2}\, \frac{\omega^2\omega'^2}{\omega-\omega'-\rmi\,0}
\int d{\bf r}''\int d{\bf r}'''\int d{\bf r}''''\int d{\bf r}^v\, 
\bfsfT^\ast({\bf r},{\bf r}'',\omega)\nonumber\\
\fl \cdot \bfsfG({\bf r}'',{\bf r}''',\omega-\rmi\, 0)\cdot
\left[\bchi({\bf r}''',{\bf r}'''',\omega-\rmi\, 0)
-\tilde{\bchi}^\ast({\bf r}'''',{\bf
  r}''',\omega'-\rmi\, 0)\right]\nonumber\\
\cdot\tilde{\bfsfG}^\ast({\bf r}^v,{\bf r}'''',\omega'-\rmi\, 0)
\cdot \tilde{\bfsfT}({\bf r}',{\bf r}^v,\omega').
\label{C.7}
\end{eqnarray}

Having obtained all contributions to the integral kernel $\bfsfM$ we are
now in a position to evaluate their sum. We start by investigating the
terms with transverse Green functions, as given by (\ref{C.5}),
(\ref{C.6}) and part of (\ref{C.4}). Taking all terms together we may
write them as
\begin{eqnarray}
\fl \mu_0\, \hbarit\int d{\bf r}''\int d{\bf r}'''\int d{\bf r}''''\, 
\bfsfT^\ast({\bf r}, {\bf r}'',\omega)\cdot\left\{\rule{0mm}{5mm}
-\omega\, \left[
\bfsfG({\bf r}'',{\bf r}''',\omega-\rmi\, 0)\right]_{T'''}\right.\nonumber\\
 \cdot\left[\bfsfI\, \delta({\bf r}'''-{\bf r}'''')-
\frac{\omega'^2}{c^2}\, \tilde{\bfsfG}^\ast({\bf r}'''',{\bf
  r}''',\omega'-\rmi\, 0)\right.\nonumber\\
\left. -\frac{\omega'^2}{c^2}\int d{\bf r}^v\, 
\tilde{\bchi}^\ast({\bf r}^v,{\bf r}''',\omega'-\rmi\, 0)\cdot
\tilde{\bfsfG}^\ast
({\bf r}'''',{\bf  r}^v,\omega'-\rmi\,0)\right]\nonumber\\
 -\omega'\left[\bfsfI\,\delta({\bf r}''-{\bf r}''')-
\frac{\omega^2}{c^2}\, \bfsfG({\bf r}'',{\bf r}''',\omega-\rmi\,0)\right.\nonumber\\
\left. -\frac{\omega^2}{c^2}\int d{\bf r}^v\, 
\bfsfG({\bf r}'',{\bf r}^v,\omega-\rmi\,0)\cdot
\bchi({\bf r}^v,{\bf r}''',\omega-\rmi\, 0)\right]\nonumber\\
\left. \cdot\left[\tilde{\bfsfG}^\ast
({\bf r}'''',{\bf r}''',\omega'-\rmi\,0)\right]_{T'''}\right\}\cdot
\tilde{\bfsfT}({\bf r}',{\bf r}'''',\omega').
\label{C.8}
\end{eqnarray}
The sets of terms that multiply the transverse Green functions are
transverse themselves, as follows from (\ref{3.11}). Since the
integral of the scalar product of a longitudinal and a transverse function
vanishes, we can replace each of these transverse Green functions by their
full Green function counterparts. Subsequently, upon using (\ref{3.11}) and
performing partial integrations we may rewrite (\ref{C.8}) in the form
\begin{eqnarray}
\fl \mu_0\,\hbarit (\omega+\omega')\int d{\bf r}''\int d{\bf r}'''
\int d{\bf  r}''''\,\bfsfT^\ast({\bf r},{\bf r}'',\omega)\cdot
\left\{\left[\bfsfG({\bf r}'',{\bf r}''',\omega-\rmi\,0)\times
\overleftarrow{\bnabla}'''\right]\times
\overleftarrow{\bnabla}''''\right\}\nonumber\\
\cdot \tilde{\bfsfG}^\ast({\bf r}'''',{\bf r}''',\omega'-\rmi\,0)\cdot
\tilde{\bfsfT}({\bf r}',{\bf r}'''',\omega').
\label{C.9}
\end{eqnarray}
An alternative form for this expression is found upon splitting it in two
terms by writing the factor $(\omega+\omega')$ as the difference of
$\omega^2/(\omega-\omega'-\rmi\,0)$ and
$\omega'^2/(\omega-\omega'-\rmi\,0)$. Subsequently, we carry out partial
integrations in the first term, while we leave the second as it
stands. Finally, we use (\ref{3.11}) to eliminate the double spatial
derivatives. We end up with a set of terms that precisely cancel
(\ref{C.7}) and the remainder of (\ref{C.4}) (i.e., the terms without the
transverse Green functions). 

Collecting the results we find that we are left with (\ref{C.3}).  The
result for $\bfsfM$ is thus quite simple:
\begin{equation}
\bfsfM({\bf r},{\bf r}',\omega,\omega')=
\bfsfI \,  \delta({\bf r}-{\bf r}')\, \delta(\omega-\omega').
\label{C.10}
\end{equation}
As a consequence, condition (\ref{C.2}) becomes:
\begin{equation}
\int d{\bf r}''  \, \bfsfu({\bf r},{\bf r}'',\omega)\cdot
\tilde{\bfsfu}^\ast({\bf r}',{\bf r}'',\omega)=
\bfsfI \,  \delta({\bf r}-{\bf r}').
\label{C.11}
\end{equation}
In other words, the tensorial integral kernel $\bfsfu({\bf r},{\bf
  r}',\omega)$ must be unitary. For convenience we choose from now on:
\begin{equation}
\bfsfu({\bf r},{\bf  r}',\omega)=\bfsfI\, \delta({\bf r}-{\bf r}')
\label{C.12}
\end{equation}
so that $\bfsfu$ is independent of the frequency and diagonal in the
spatial variables. A different choice for $\bfsfu$ leads to a unitarily
equivalent form of the diagonalizing operator ${\bf C}({\bf r},\omega)$, as
follows from (\ref{3.2}) with (\ref{B.15})--(\ref{B.18}). Upon inserting
(\ref{C.12}) in (\ref{B.15})--(\ref{B.18}) we finally arrive at 
expressions (\ref{3.7})--(\ref{3.10}) of the main text.

As a final check of the expressions for $\bfsff_i$ we may verify that the
commutator $[{\bf C}({\bf r},\omega),{\bf C}({\bf r}',\omega')]$ vanishes
for all position and frequency arguments. To that end we have to check
whether the  condition
\begin{eqnarray}
\fl \rmi\, \hbarit\int d{\bf r}''\, \left[
\bfsff_1({\bf r},{\bf r}'',\omega) 
\cdot \tilde{\bfsff}_2({\bf r}',{\bf r}'',\omega')-
\bfsff_2({\bf r},{\bf r}'',\omega) 
\cdot \tilde{\bfsff}_1({\bf r}',{\bf r}'',\omega')\right] \nonumber\\
\fl +\int d{\bf r}''\int_0^\infty d\omega''\, 
\left[
\bfsff_3({\bf r},{\bf r}'',\omega,\omega'') 
\cdot \tilde{\bfsff}_4({\bf r}',{\bf r}'',\omega',\omega'')\right.\nonumber\\
\left.-\bfsff_4({\bf r},{\bf r}'',\omega,\omega'') 
\cdot \tilde{\bfsff}_3({\bf r}',{\bf
  r}'',\omega',\omega'')\right]=0
\label{C.13}
\end{eqnarray}
is satisfied. Along similar lines as above one verifies that this is
indeed true.

\end{document}